\documentclass{jkas}

\usepackage{caption}
\usepackage{subcaption}

\usepackage[export]{adjustbox}

\def\year{2016} 
\def\month{September} 
\def\volume{00} 
\def\issue{0} 
\def\beginpage{1} 
\def\endpage{99} 
\def\received{---} 
\def\accepted{---} 

\date{Received \received ; accepted \accepted}

\def\lim{{\rm lim}}

\title{
     Proto-type installation of a double-station
     system for the optical-video-detection and orbital 
     characterisation of a meteor/fireball in South Korea
}

\author[1]{Tobias C. Hinse}
\author[2]{Woo-Kyum Kim}
\author[1]{Sang-Hyeon Ahn}
\author[2,3]{~~~~~~~~~~~~~~~~~~~~~~~~~~~~~~~~~~~~~~~~~~~~~~~~~~~~~~~~~~~~~~~~~~~~~~~~~~~~~~~~~Jae-Keun Lee}
\author[2,4]{Jun-Hyeong Park}
\author[2,5]{Young-Woo Lee}
\author[2,6]{Woo-Jung Jeong}
\author[2,7]{Sang-Min Woo}

\affil[1]{Korea Astronomy \& Space Science Institute, Daejon 305-348, Republic of Korea.
\email{tchinse@gmail.com}}
\affil[2]{Daejeon Science Highschool, 305-338 Daejeon, Republic of Korea.}
\affil[3]{Seoul National University, College of Engineering, Department of Energy Resources Engineering, Seoul, Republic of Korea}
\affil[4]{Hanyang University, College of Engineering, Department of Electronic Engineering, Seoul, Republic of Korea}

\affil[5]{Seoul National University, College of Agriculture and Life Sciences, Department of Forest Sciences, Seoul, Republic of Korea}
\affil[6]{Hanyang University, College of Natural Sciences, Department of Life Science, Seoul, Republic of Korea}
\affil[7]{Korea Advanced Institute of Science \& Technology, Daejeon, Republic of Korea}

\def\corrauthor{
Tobias C. Hinse
}

\def\runningauthor{
Hinse et al.

}

\def\runningtitle{
    Optical Video-Meteor Detection in South Korea
}

\def\keywords{
meteor detection -- meteor showers -- orbit determination -- astrometry -- comets -- asteroids
}

\def\abstracttext{
We give a detailed description of the installation and operation of
a double-station meteor detection system which formed part of a 
research \& education project between Korea Astronomy  
Space Science Institute and Daejeon Science Highschool. A total of 
six light-sensitive CCD cameras were installed with three cameras at 
SOAO and three cameras at BOAO observatory. A double-station 
observation of a meteor event enables the determination of the 
three-dimensional orbit in space. This project was 
initiated in response to the Jinju fireball event in March 2014. The 
cameras were installed in October/November 2014. The two stations are 
identical in hardware as well as software. Each station 
employes sensitive ``Watec-902H2'' cameras in combination with 
relatively fast f/1.2 lenses. Various fields of views were used for 
measuring differences in detection rates of meteor events. We employed 
the SonotaCo UFO software suite for meteor detection and their 
subsequent analysis. The system setup as well as installation/operation 
experience is described and first results are presented. We also give a 
brief overview of historic as well as recent meteor (fall) detections 
in South Korea. For more information please consult 
http://meteor.kasi.re.kr}

\begin{document}
\jkashead 


\section{Introduction \label{sec:intro}}

The dynamical evolution and inventory of the solar system can be understood by studying the interaction between planets on the largest and meteoroids on the smallest scale. The detection of atmospheric meteorite events can be used to identify a group of meteors that appear to originate from a common point on the sky - the radiant. Each group produce annual meteor 
showers \citep{jen2017} as a consequence of the orbital motion of our home planet around the Sun. From accurate timing and astrometric triangulation measurements the trajectory and velocity vector of the meteoroid can be determined and its point of appereance on the sky (radiant)) located. At the time of writing (June 2017) the IAU Meteor Data Center\footnote{URL: https://www.ta3.sk/IAUC22DB/MDC2007/index.php} lists a total of 726 meteor showers of which 112 are classified as established showers. Once a meteor shower group has been identified numerical integration of a cloud of particles can be followed over time to establish a link between the shower properties and a solar system parent body. The most famouse meteor shower is the Orionids shower with peak count (zenithal hourly rate or ZHR) rate in late October. The meteoroid stream causing each meteor shower has been formed of debris emitted regularly from its parent body's close approaches to the Sun. Thus, meteor streams have usually the similar orbital characteristics to their parent bodies. Most of parent bodies for meteor showers are comets, while some of them are asteroids.

The identification of meteor showers is not trivial and relies on several meteor detection techniques. The most wide-spread technique is the detection by means of video-recording equipment. The technological development and steep decrease in cost over the past 10 - 15 years had the result of an increase interest in meteor detections supported mostly within the amateur astronomy community. Only meteor events observed from a minimum of two independent detectors can be used for the determination of the meteoroid orbit. The optimal baseline of the two observing stations is on the order of 100 km. 

In this paper we give a detailed description of designing and developing a double-station meteor detection system installed at the Korean-based SOAO and BOAO observatories. Several advantages are to be considered. The distance between the two observatories is nearly perfect amounting to just under 100 km. Furthermore, the infrastructure allows the installation of a fixed observing setup with continuous and stable power-supply. The most important factor is their height above sea-level and remote location from nearby cities in order to take advantage of a low sky background brightness.

Several multi-station meteor detection initiatives exist in various countries. The two main detection techniques are based on video (optical) and radar \citep{fleet2015} observations. Another technique, which is less practiced, is based on infra-sound measurements \citep{silber2009, silber2014}. 

In the following we list a number of collaboration specialised on the detection of meteors. The UKMON\footnote{UK Meteor Observation Network} initiative \citep{ukmon2014} which collects data from 15 cameras mainly located in Southern England. The Armagh Observatory/Bangor single and double-station network in Northern Ireland, UK \citep{armaghmeteor2007}. The Spanish Meteor Network (SPMN) \citep{spmn2004,spmn2007} which uses all-sky cameras for meteor and fireball detections at four stations and is operational since 2004. The Southern Ontario Meteor Network operates an automated all-sky camera network \citep{asgard2010} in Canada and makes use of self-developed data processing software (ASGARD). The Sonotaco \citep{sonotaco2009, kanamori2009} network of over 100 cameras located in mid- and south Japan and the more recent CAMS (Cameras for Allsky Meteor Surveillance) system \citep{CAMS1} operating a 3-station network of 20 cameras each and located in California, USA. In France the French Fireball Network (FRIPON, Fireball Recovery and Interplanetary Observations Network, \citet{atreya2011, colas2015}) is probably the most ambitious network at current time and aims for the precise detection of meteors and fireballs with a total of 100 all-sky stations each separated by about 100 km.

Several meteor detection software packages exist. One particular promising open-source project was recently presented by \citet{vida2016} and utilizes the relatively 
inexpensive and small-format Raspberry PI (2 and 3) hardware platform. Their idea seem fruitful solving the onsite installation problem effectively for a multi-camera (all-sky) setup. More widespread software for meteor detection is {\sc MetRec} \citep{molau1994} and {\sc MeteorScan} \citep{gural2008}. A user-friendly implementation of meteor detection can be found in the {\sc UFO} suite of software packages (mainly {\sc UFOCapture}) \citep{sonotaco2009}. The latter software is employed in the present description of a meteor detection network. Finally the Ontario network of fireball detection have developed {\sc ASGARD} as their own detection software \citep{asgard2010}. A performance comparison between {\sc ASGARD} and {\sc UFOCapture} was carried out by \cite{blaauw2012}.

This paper is structured as follows. In section \ref{sec:meteorfalls} we give a brief review of recent and historic meteor falls in Korea. Section \ref{sec:instru} gives a detailed description of the instrumental setup encompassing details on the camera, lens, mounting 
and peripheral computing equipment. Section \ref{sec:dataproc} gives a overview description of the various branches within the {\sc UFO} software package including a discussion of solving the time synchronisation problem. First results obtained during the period November 2014 to March 2015 are presented in section \ref{sec:results} and we summarize and provide an outlook for future upgrades and solutions to current problems in section{sec:summary}.

It is the hope of the authors that this paper might also be of interest to the Korean 
amateur astronomy community and inspires people from the general public or public astronomical / meteorological observatories located across the country (preferrably in rural areas on mountain tops) to setup their own (in-expensive) meteor detection system and actively contribute with accurate data. Maybe KASI can serve as a national hub to collect all the gathered data in the future. For assistance please contact the lead author.

\begin{figure}
\centering
{\includegraphics[width=0.47\textwidth]{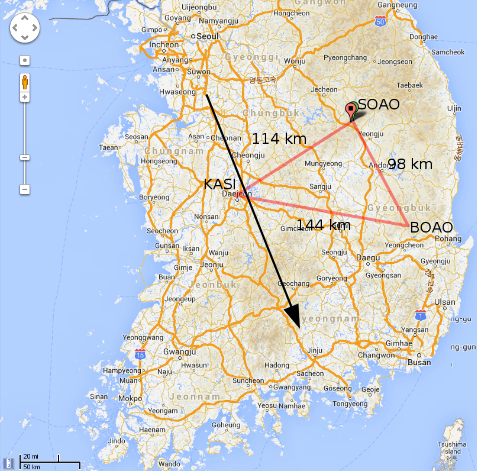}}
\caption{Geographic location of BOAO and SOAO observatories in relation to Daejeon (KASI). The baseline distance between the two observatories is around 100km. The black arrow gives a rough projected path of the 2014 Jinju fireball. \emph{See online version for colors}.}
\label{soao_boao_jinjupath}
\end{figure}

\section{Historic and recent meteor falls in Korea \label{sec:meteorfalls}}

\subsection{Meteor records in the past}

Meteors and meteorites were regarded as bad omen to many cultures in the premodern times of human existence. 
Thus, they have long been an attractive phenomena in history, as is also the case in the Korean history. 

The meteor records in the Korean chronicles dates back to two thousand years ago. The first records
dates back to 104 CE of the Silla dynasty: ``Many stars fell like a rain, but they did not reach the
ground'', which is certainly a record of meteor outburst. Most of the astronomical observations in the
era of the Three Kingdoms (57 BCE to 668 CE) and the era of the Unified Silla dynasty (669 CE to 935 CE) 
are listed in the Chronicle of the Three Kingdoms ({\it Samguk-sagi}) edited by \citet{kim1145}.
According to the written archived chronicle, the Silla dynasty (54 BCE to 918 CE) left observational 
records of 9 meteor outbursts and 21 meteor/fireballs; Goguryeo (37 BCE to 668 CE) left 2 outbursts and 1
meteor/fireball; and Baekje (18 BCE to 660 CE) did 2 outbursts and 3 meteor/fireballs. The records of
astronomical phenomena in the Goryeo dynasty (918 CE to 1392 CE) are preserved in both the
Chronicles of the Goryeo dynasty ({\it Goryeosa}) written by \citet{kim1451} and the Simplified
Chronology of the Goryeo dynasty ({\it Goryeosa-Joryo}) written by \citet{kim1452}, in which we can find
729 records of meteors \citep{ahn2003,ahn2005}.

Analysis of the meteor/fireball records during the Goryeo dynasty proved the existence of presently
conspicuous meteor shower such as the Perseids, the Leonids, and the eta Aquarids/Orionids pair
formed by Halley's comet \citep{ahn2003}. The Goryeo records are combined with those in the Chronicles
of Song China to show the existence of the annual variations of sporadic meteors, which is ascribed
to the inclination of the Earth's rotation axis to the orbital plane. In addition, there were at least
two conspicuous meteor showers such the Perseids and the Leonids. The regression rate of the
Leonids is measured to be approximately 1.5 days per century, which agrees with the modern
estimates \citep{ahn2005}.

The meteor records of the Joseon dynasty (1392 CE to 1910 CE) are preserved in the Royal Chronicles of 
the Joseon Dynasty ({\it Joseon-wangjo-sillok}) and the Daily Records of Royal Secretariat of the Joseon 
Dynasty ({\it Seungjungwon-Ilgi}). We find more than 3,500 records of meteors, mostly fireballs 
\citep{ahn2005}. The former spans from 1392 CE to 1910 CE, while the latter spans from 1623 CE to 1910 CE. 
Although the durations encompassed by the two data sets are different, the numbers of meteor/fireball 
records are similar to each other, i.e. approximately 3,500 records, respectively. Analyses for these data 
sets show that there were persistent meteor showers as had been seen in the Goryeo dynasty 
\citep{ahn2005}.

There are also records of meteor outbursts written in the Korean historical chronicles. Adding the 
meteor outbursts from world-wide historical archives to the Korean data, we can see the persistency of 
several meteor outbursts such as the Lyrids, the Perseids, the Leonids, and the eta Aquarids/Orionids 
pair during the last one or two millennia. Through these data we can find some hint on the long-term 
evolution of their orbits. For example, the Leonids show abrupt emergence during the 9th century. The 
meteor streams such as the Geminids and the Quadrantids seem to emerge in the relatively recent times. 
Several meteor streams shows the evidence of their orbital variation possibly due to the precession of 
their orbits \citep{ahn2015, ahn2016}.

\subsection{Historic meteorites falls}
There are also records of meteorite falls and their recoveries. During the three kingdoms era, there are 
only records of fireballs called {\it Cheongu-seong} meaning heavenly dog star, namely big fireballs 
disappearing beyond the landscape. However, in general, there is no description in the records on the 
recovery of meteorites. There are approximately seven highly reliable records of meteorite 
falls in the Chronicles of the Goryeo dynasty. One certain case is that of 1070 CE, which says, 
``in the Kyeongja day of the 1st month of the 24th reign year of emperor Munjong of the Goryeo dynasty, 
in Daegu, a star fell to the ground and became a stone''. Another one in 1294 CE describes certainly the 
conspicuous appearance of the meteorite: ``In the 3rd month of the 25th reign year of the King 
Chungnyeol-wang, in Nisan-hyeon, a meteorite fell. Its material seemed to be a jade and its shape 
resembled a chicken egg''.

According to the royal Chronicles, there were also ten records indicating certainly meteorite falls 
during the Joseon dynasty (1392 CE to 1910 CE). One example is a meteorite fall in Hamgil-do Yongjin-hyeon 
(presently Hamgyeongnam-do Muncheon-gun): ``On the 16th day of the 2nd month of the 2nd reign year of 
King Munjong (1452 CE), a fireball fell to the ground, on which a pit was formed with its circumference of 
31.5 feet''. That means the meteorite crater has a diameter of approximately 10 feet or 2-3 meters. Another 
example, is that of the King Seongjong era: ``On the 1st day of the 4th month in the 23rd reign year of King 
Seongjong (1492 CE), when there was thunder-storm and heavy rain, a meteorite fell in Jinju. That entered 
into the ground by one feet in depth. A brave soldier Kang Kyeson excavated and found the meteorite, whose 
color is that of a noeseol and its shape looks like a bokryoeng. When being scratched with a nail, its powder 
fell off''. Here {\it noeseol} is a sort of mushroom growing on the root of bamboo, whose surface is black and 
the interior is white resembling chestnut. {\it Bokryeong} is also a sort of mushroom growing on the root of 
a pine-tree, whose surface is brown in color and has many wrinkles resembling a lump as big as a ball. Thus, 
the meteorite must have been a chondrite meteorite having its fusion crust.

\subsection{Recent meteorite falls in Korea}

In modern times before the fall of Jinju meteorite in 2014, we have four meteorite records registered in 
Catalogue of Meteorites edited by \cite{grady2000}: Ungok (Unkoku in Japanese) meteorite weighs 1 kg and 
is a chondrite found on Sep 7, 1924, Okgyei (Gyokukei in Japanese) meteorite weighs 1.32 kg and is also a 
chondrite found in March 7, 1930, Sobaek (Shohaku in Japanese) meteorite weighs 0.101 kg and is an iron 
meteorite, and Duwon (Duwun in Japanese) meteorite. All of them were either falls or finds during the period 
of Japanese imperialism, and now only one is preserved. That meteorite is Duwon meteorite, which was found 
in Seongdu-ri 186-5, Duwon-myeon, Goheung-gun, Jeollanam-do, Republic of Korea, on 23 November 1943 15:47 
with lightening and sonic boom by a local Japanese school principal. It was taken to the discoverer's home 
land Japan after Korean independence, and housed in the National Science Museum of Japan. Later, in 1999, 
it was turned back to Korea, and now it is housed in the geological museum of Korea Institute of Geoscience 
and Mineral Resources in Daejeon. The meteorite weighs 2.117 kg was classified as L6-type ordinary 
chondrite \citep{ahn2002, choi2002}.

\begin{figure*}
\centering
{\includegraphics[width=0.49\textwidth]{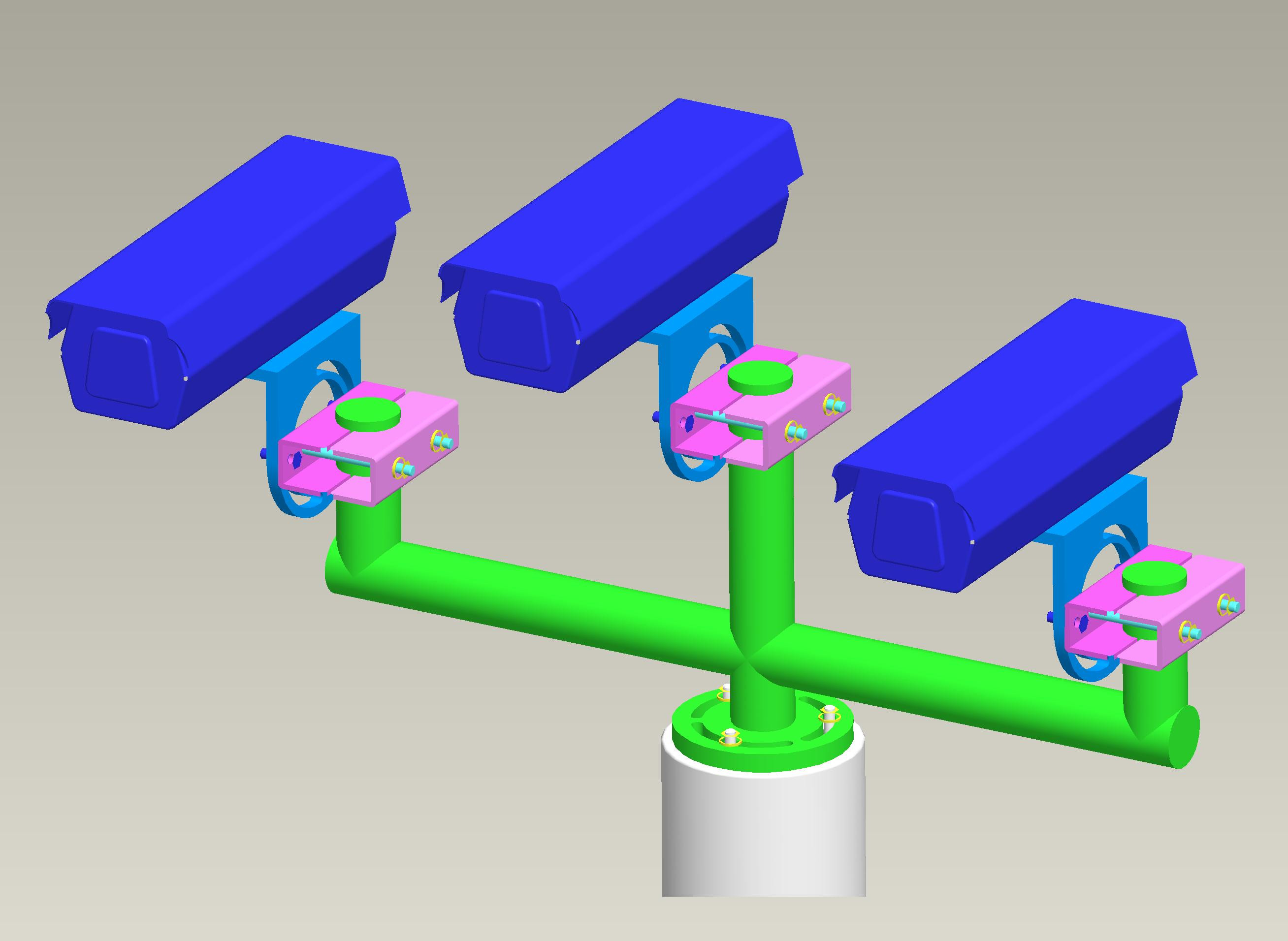}}
{\includegraphics[width=0.4885\textwidth]{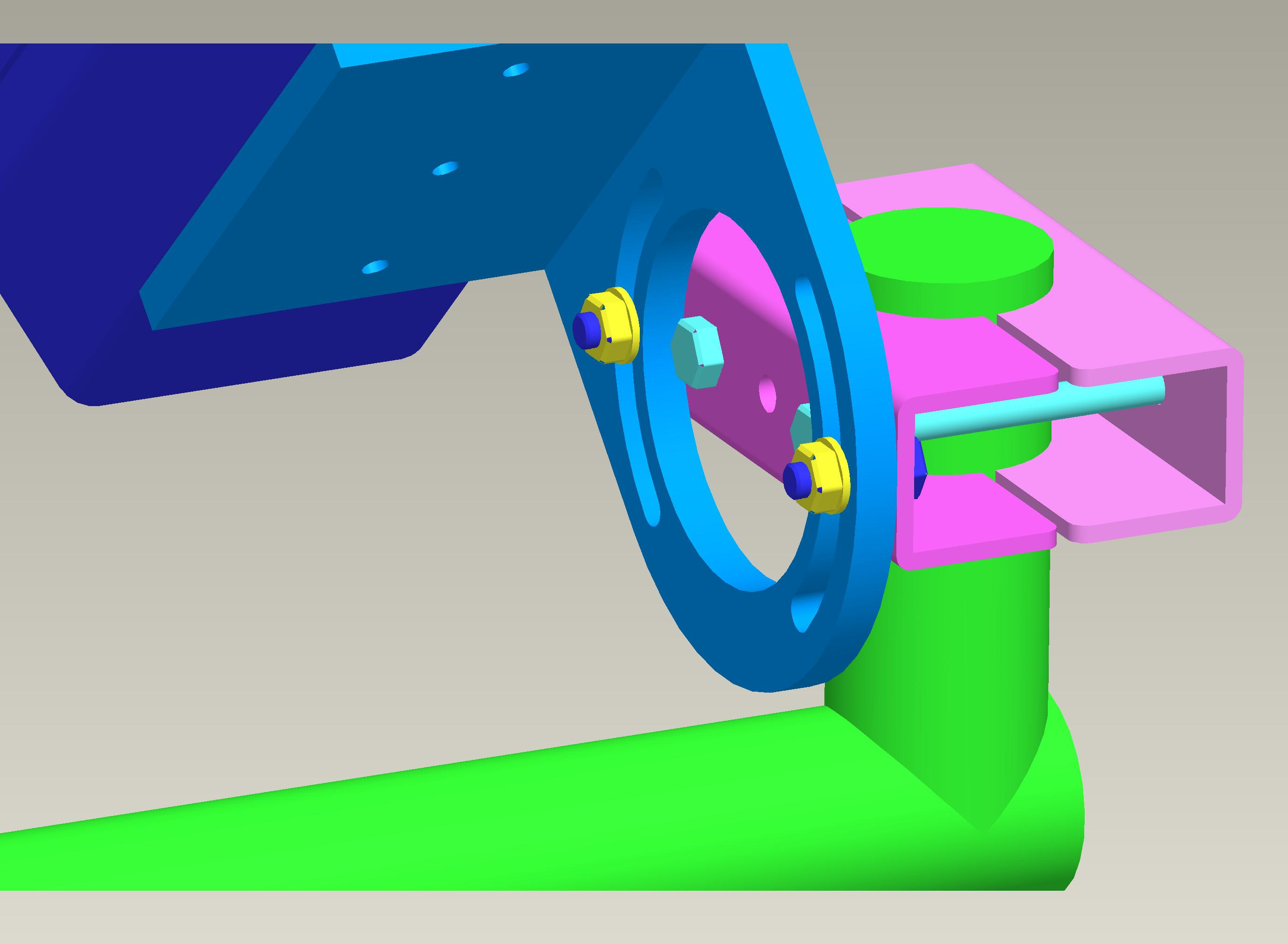}}
\caption{Result of initital phase 3D design study of camera mount and setup. Special attention was paid towards the correct dimensioning to ensure rigidity and stability in order to resist disturbing factors from the environment (strong wind mainly). The mount consist of two main parts: a lower cylindrical part and a top three-fork part on which the cameras are mounted. The top part is flanged-mounted to the lower part via three bolts and can be rotated around the vertical axis nearly 360 degrees. The benefit of this feature is to change the azimuth angle for all three cameras at \emph{once} for future experimentation when different sky regions are needed to be probed. This is possible by allowance of a complete elongated grove in the base flange of the three-fork part with the outer rim connected at three contact points to the main metal structure. This design also ensure ease of transportation when dismantling the mount into single parts. \emph{See online version for colors}.}
\label{3ddesign}
\end{figure*}

More recently, on March 9, 2014, 20:04 (KST) a fireball was observed travelling south-bound more than 300 km 
from the city of Suwon, and its sonic boom was heard at the southern part of the Korean peninsula 
(see Fig.~\ref{soao_boao_jinjupath}). The fireball was imaged by a number of blackbox-cameras installed 
within the car, but those image data and time recordings were not accurate enough to be analyzed to provide 
detailed orbital information of the Jinju meteor event.

To make it worse, there was no national meteor surveillance system available at that time in Korea that were 
developed by professional astronomers. Four meteorites of 9 kg, 4.1 kg, 0.4 kg, and 20.5 kg each were recovered 
in a few days after the fireball at Jinju area. Those meteorites were simply analyzed scientifically to be 
classified as H5 ordinary chondrite based on their petrological characteristics and chemical and oxygen isotopic 
composition in addition a number of geochemical analyses \citep{choi2014,nagao2015,choi2015,goh2016}
were carried out. 

In the same year as the occurance of the Jinju fireball, the existence of an iron meteorite was reported 
(Jwa, Y.-J., interview, Yonhap News, July 2, 2014). The meteorite was found in 1970s on a field in 
Miwon-myeon, Cheongju, Republic of Korea. The meteorite weighs 2.008 kg, and analyses prove that the meteorite 
is an iron meteorite with a relatively low Ni abundance.

\section{Instrumental setup at ground-stations \label{sec:instru}}

The two newly installed meteor detection stations\footnote{SOAO - IAU observatory code: 
345; BOAO - IAU observatory code: 344} SOAO and BOAO are nearly identical and  employes fixed-oriented, 
identical setup of mounts, applied electronics, signal transportation, camera enclosures, camera and optical lens equipment. Initially we thought 
about different mounting designs in order to ensure stability and rigidity against possible strong sheer-winds on remote mountain tops. Fig.~\ref{3ddesign} shows details of the final 3D mount arrangement of a single observing station and Fig.~\ref{boaosoaocams} shows the final/current installation of the meteor detection system at SOAO/BOAO 
observatory. 

The actual video processing uses different computer hardware and various analog-to-digital (AD) video-signal conversion equipment (details to be described in a forthcoming section). To ensure ease of post-processing of idential data products we decided to use the SonotaCo UFO software package for data acquisition/detection and analysis (see details in forthcoming sections). In overall the system has been in a reasonable good autonomous working operation since October/November 2014 with minor problems experienced only on the computer end of the setup. The initial setup is a novum to be installed in Korea and touches very close to a professionally installed meteor detection system. However, the setup is far from perfect and the present installation serves as a proto-type benchmark test-setup to gain experience in research \& education, project management, installation, technology testing, working operation and data processing and analysis of meteor detection in the Republic of Korea.

It is no secret that inspiration and the realisation of this project very much relied on the unexpected event of the 2014 Jinju fireball, the installation of a similar camera system at the Armagh Observatory \citep{armaghmeteor2007} and the detailed description of a US-lead meteor detection project \citep{CAMS1}.

\subsection{Used hardware \label{sec:hardware}}

\begin{figure*}
\centerline{{\includegraphics[scale=0.3]{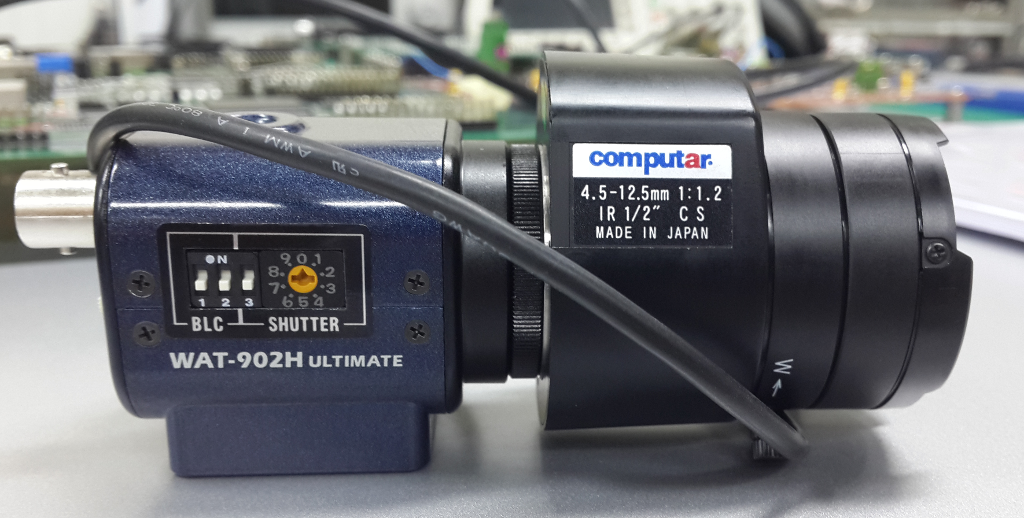}}}
\caption{Installed camera and lens system. The lens is a vari-focal lens enabling different field of views at changing resolution. The cable enables the auto-iris function of the 
lens and connects to the camera on the rear-side. Often the model designation ``902-H2'' is used for the official ``902-H Ultimate'' model name. \emph{See online version for colors}.}
\label{camandlens}
\end{figure*}

\begin{figure}
\centering
{\includegraphics[width=0.47\textwidth]{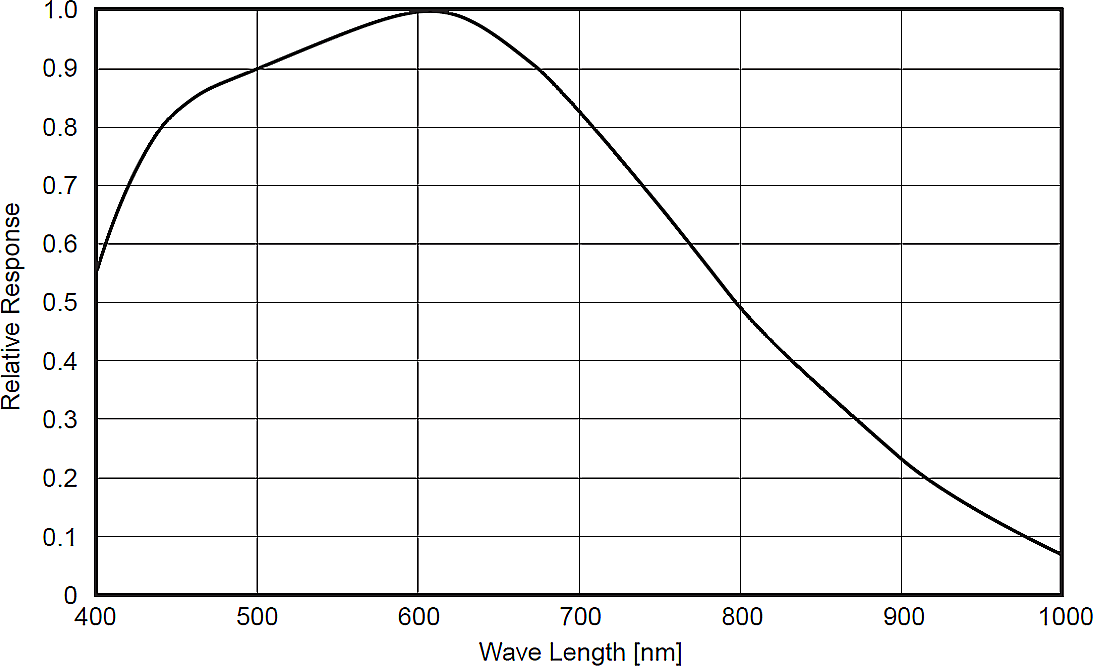}}
\caption{Optical transmission curve for the Watec-902H2 camera. Additional optical systems will degrade the light transmission efficiency. \emph{Image credit: Courtesy of Watec corporation}.}
\label{specresponse}
\end{figure}

\begin{figure}
\centering{\includegraphics[width=0.37\textwidth]{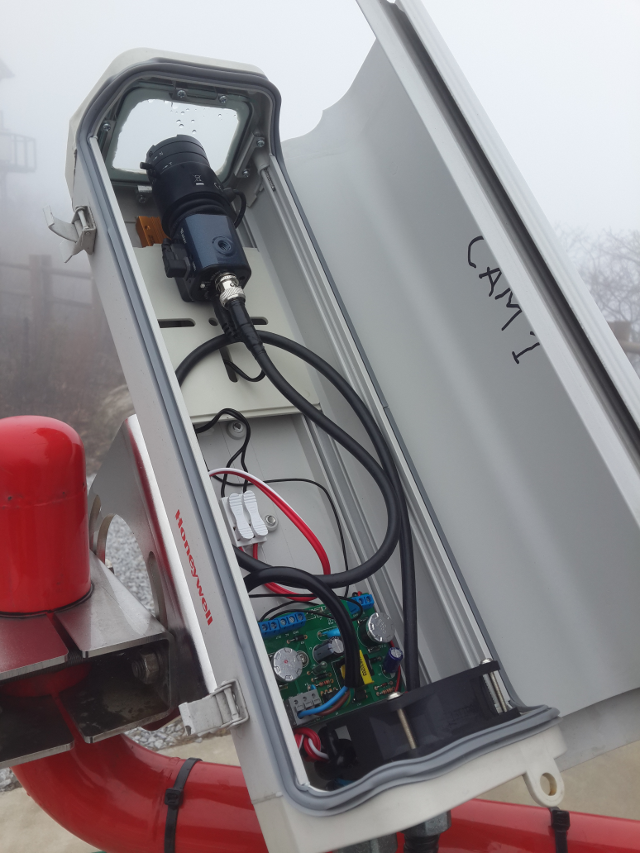}}
\caption{Detailed view of the interior of the protective camera housing of SOAO camera 1. Both the coaxial video-signal and 12V camera power cable (white/red) are visible. Additional electronics controlls the housing fan (mounted on the rear-end and heater (placed under the camera lens beneath the housing front glass). \emph{See online version for colors}.}
\label{insidehouse}
\end{figure}

The cameras used are of type Watec-902H2 as shown in Fig.~\ref{camandlens}. Each camera is securely embedded into a standard CCTV camera housing (Honeywell, model GHG-140SHBA-220) featuring a fan/heater (separate 110VAC/220VAC power supply needed) to prevent dew 
condensation or over-heating on the inside (cf. Fig.~\ref{insidehouse}). This ensured a tolerable working environment of the camera and related optics. The cooling aspect is important since the camera itself generates a significant amount of heat during operation hours (though during winther months this imposes no thread to normal operation of the camera). The heating element is installed right beneath the front glass. 

The housing is waterproof and rated against the IP65 waterproof-tested industrial standard and proofed itself to be sufficient given the harsh weather conditions during the Korean monsoon and winther season. However, since the housings are mounted to point upward toward the sky (rather than top-down as in usual urban-surveillance installation sites like parking space or building entrances) the front assembly screw-holes needed to be silicone-sealed in order to prevent water intrusion. Failure to do so will certainly cause the housing to be flooded during the rain season in the summer months. 

Efforts to reduce reflection effects were omitted due to budget limitations. It was decided not to carry out special coating treatment (anti-reflection) of the housing front glass. We estimate that on average a 75\% transmission in the wave-length range 480 - 620 nm is reasonable for the chosen camera elevations and related optics \citep[their fig. 3]{CAMS1}.

In the future we plan to carry out detailed transmission response measurements considering various circumstances in order to explore the potential to increase the optical transmission rate (which again will increase the sensitivity and hence the meteor detection rate). We refer to a forthcoming section for details on the optical properties of the lens.

\subsection{Technical camera description\label{sec:cameratech}}
The Watec-902H2\footnote{http://www.wateccameras.com/product\_documents/902H2} (EIA type) camera is a un-intensified low-light level monochrome camera and available as an ``off-the-shelf''product at a reasonable price. The camera is originally designed for CCTV surveillance purpose in low-light areas. Due to its high sensitivity (advertised as 100 microlux at f/1.4) it is widely used in the professional as well as amateur meteor detection community around the world. Especially in the area of stationary long-duration sky-monitoring surveys of variable stellar objects. Compared to 
the Watec-902H and Watec-902A camera models the Watec-902H2 model is 2.5 times more sensitive in the visible and 3 times more sensitive in the near-infrared wavelength range. From experience made in the international community the camera proofs itself to be more reliable compared to alternative (though more sensitive) light-intensified cameras.

Each camera has a 1/2" format interline transfer CCD sensor (type SONY ICX428AL) that utilises SONY Exview Hole Accumulation Diode (HAD) technology to increase sensitivity and compared to other CCD's, offer improved quantum efficiency, reduced smear and in overall reduce dark-current background noise. The image scanning system is standard 1/2 interlaced which means that half of the image is shown followed by the second half 1/60s later and produces a standard analog composite NTSC video-signal (1V peak-to-peak, 75 $\Omega$, unbalanced). The number of total pixels is 811 (h)orizontal x 508 (v)ertical of which 768 (h) x 494 (v) are used for image production. However, the AD video-signal converters produces a final output image of 640 (h) x 480 (v) pixels at 29.97 frames per second (NTSC). The physical CCD pixel size is 8.4 $\mu$m (h) and 9.8 $\mu$m (v). In Fig.~\ref{specresponse} we show the spectral response of the camera with more than 75\% transmission efficiency in the wave-length range 450 - 720 nm.

\begin{figure*}
\centering
\includegraphics[width=0.34\textwidth]{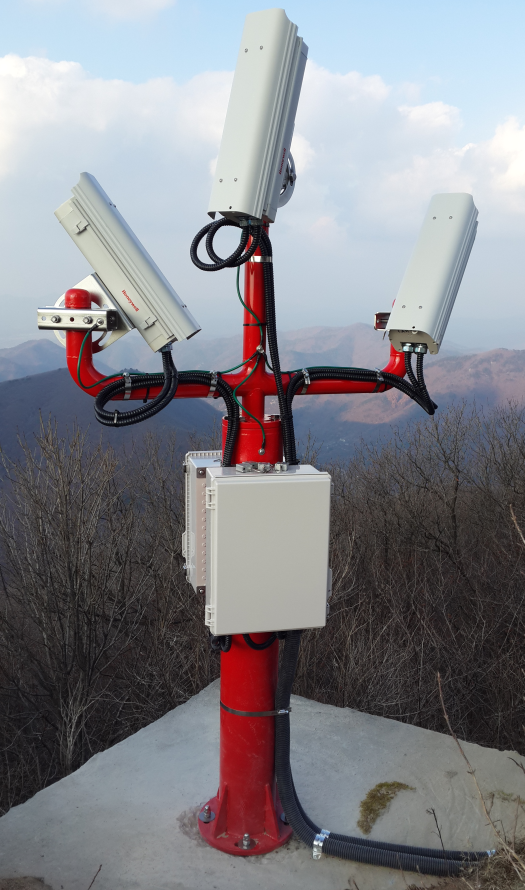}
\vspace{2mm}
\includegraphics[width=0.34\textwidth]{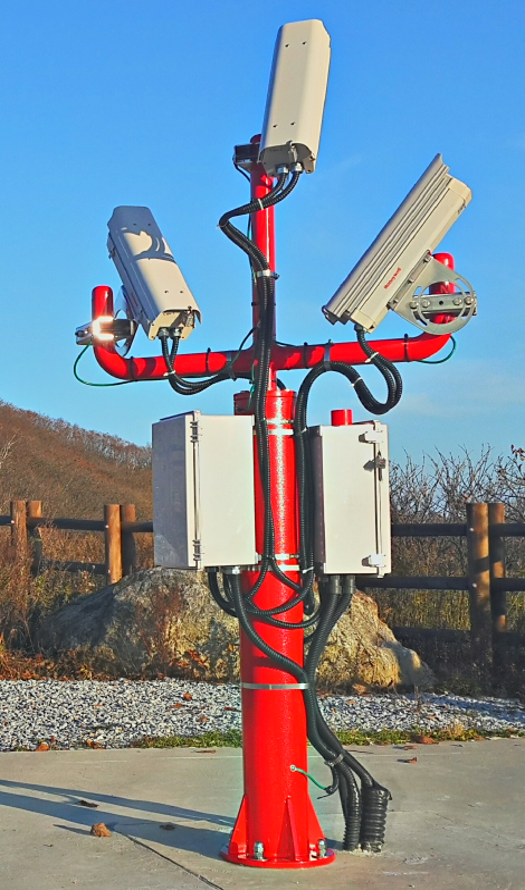}
\caption{\emph{Left picture:} Camera site at BOAO observatory. \emph{Right picture:} Camera site at SOAO observatory. All cameras are mounted in a fixed position. Camera number 1,2 and 3 for the BOAO site are from right to left. For the SOAO site the camera numbering is from left to right. We refer to Table \ref{astrodetails} for details on orientation and field of view. The two gray boxes houses the video-cable and power-supply equipment. \emph{See online version for colors}.}
\label{boaosoaocams}
\end{figure*}

\subsection{Physical camera operation settings\label{sec:camsettings}}

The camera has several ``on-board'' adjustment settings that directly affect the sensitivity and/or image quality. 

We opt to operate the camera shutter in off-mode implying that the shutter time is 1/60s which is the minimal opening time for the used model providing a time resolution of the meteor trail of 0.0167 seconds. For meteor detection a short shutter time is preferred due to a relatively high pre-atmospheric entry speed. 

The two switches for back-light compensation (BLC) were set both to the default ``off'' position (cf. Fig.~\ref{camandlens}) implying that BLC is applied based on the whole CCD area. For meteor detection purpose BLC will almost never be used since contrast enhancement is never an issue on dark nights. The gamma correction switch (located on the read-end of camera) was set to ``off'' ($\gamma = 1$). 

The gain control was set to manual gain control (MGC) allowing to set the gain manually between 5 to 50 dB. On a clear and dark night we experimentally adjusted the camera gain and 
settled on a subjective trade-off between a sufficient low background noise level and a maximum number of background stars visible. However, we did no systematic quantitative experimentation to find the optimum gain setting. From a technical/scientific point of view the trade-off is between photometric sensitivity and astrometric accuracy. When performing manual gain control setting we were guided by the latter (maximising the number of stars visible in the respective field of view) to ensure useful astrometric measurements.

We judge that the limiting magnitude for meteors is around +5.0 and for background stars around +6.0 mag. For the cameras with a wide field of view those limits may be lower (see forthcoming section). For later photometric calibration studies we have frequently obtained dark frames and flat fields for each camera. As a result from astrometric measurements we 
find that each CCD chip exhibit a few hot pixels.

\subsection{Optical lens properties and field of view\label{sec:lens}}

The sensitivity or meteor detection rate can be increased by chosing a suitable ``fast'' optical lens. Further the astrometric accuracy of meteor trails depends on the CCD detector size and lens field of view. In this section we will address these issues. 

Among various lens model the Computar f/0.8 lens is the most popular lens in the international meteor-detection community due to its relatively large aperture\footnote{however, it also introduces larger optical distortion}. However, such lenses are not produced anymore for commerical purpose due to an increase in CCD chip sensitivity of modern CCD-based video 
cameras.

We have settled to equip the cameras with six identical lenses manufactured by Computar and acquired via Ebay. The lens is specifically applicable to 1/2" format CCD sensors and therefore well-suited for the Watec-902H2 camera model. The lens is of vari-focual type and has a f-stop number of f/1.2. This lens mounts directly onto to the camera via a CS type thread. The camera CS mount is back-focus adjustable for fine-tuning of lens-focus setting. A C-thread lens model would also be compatible with the Watec camera via a CS-C mount adapter.

\begin{table*}
\begin{center}
\centering
\caption{Altitude (alt.) and azimuth (az.) orientations (J2000.0 topocentric equatorial) of each camera at SOAO and BOAO stations determined from astrometry (UFOAnalyzerV2). The vertical field of views (fov-v) are calculated values as they are not provided by UFOAnalyzerV2. Parameters $k_2,k_3$ and $k_4$ are the lens aberration distortion correction parameters of 2nd, 3rd and 4th order, respectively. Camera altitude and azimuth and rotation angle are for the field of view (fov) center. $\Delta$ is the error between reference star position and center of point-spread-function (PSF) of detected star (masked) on the CCD. ``M-S'' stands for mask-star. The numbers are average (avr.) values based on five randomly selected events for which astrometric calibration was performed.}
\begin{tabular}{ccccccc} 
\hline
                                & SOAOCam1                                   & SOAOCam2 & SOAOCam3 & BOAOCam1 & BOAOCam2 & BOAOCam3 \\
\hline
\hline
\multicolumn{7}{c}{average quantities obtained from UAV2 for five random events:}                                                   \\
avr. camera alt. (deg.)     & 48.7710                                    & 56.6743  & 52.6044  & 48.3043  & 73.2533  & 51.6526  \\
avr. camera az. (deg.)      & 88.5886                                    & 132.0191 & 214.2120 & 39.4615  & 59.6645  & 281.6092 \\
avr. camera rotation (deg.)     & 0.3560                                     & 2.2832   & -0.3446  & 1.8542   & 11.9106  & -0.5427  \\
avr. $k_2$                                &-0.0232                                     & -0.0173  & -0.0606  & 0.0000   & -0.0267  & -0.0396  \\
avr. $k_3$                                & 0.0245                                     & 0.0246   & 0.0254   & 0.0051   & 0.0173   & -0.0019  \\
avr. $k_4$                                &-0.0133                                     & -0.0154  & -0.0311  &-0.0045   & -0.0018  & -0.0232  \\
avr. fov-H (deg.)                         & 46.4809                                    & 45.7653  & 79.3843  & 42.3719  & 42.3554  & 82.9797  \\
avr. $\Delta$ (pix.)                      & 0.241                                      & 0.177    & 0.237    & 0.439    & 0.280    & 0.459    \\
avr. $\Delta$ (deg.)                      & 0.018                                      & 0.013    & 0.029    & 0.029    & 0.018    & 0.058    \\
avr. num. of masks                        & 7128                                       & 8614     & 2353     & 1121     & 1517     & 1384     \\
avr. num. of M-S links                    & 213                                        & 102      & 69       & 103      & 97       & 96       \\
\multicolumn{7}{c}{derived quantities:}                                                                                             \\
fov-V (deg.)                              & 34.8607                                    & 34.3240  & 59.5382  & 31.7789  & 31.7666  & 62.2348  \\
lens focal length (mm)                    & 6.26                                       & 6.37     & 3.24     & 6.94     & 6.94     & 3.04     \\
CCD plate scale (deg./pix.)               & 0.0726                                     & 0.0715   & 0.1240   & 0.0662   & 0.0662   & 0.129    \\
CCD plate scale ('/pix.)                  & 4.36                                       & 4.29     & 7.44     & 3.97     & 3.97     & 7.74     \\
\hline
\end{tabular}
\label{astrodetails}
\end{center}
\end{table*}

The focal length of the lens is manually adjustable between 4.5 to 12.5 mm and is directly related to the size of the field of view. We have chosen to set the lenses to various focal lengths producing various field of views in order to asses the difference in sensitivity and hence detection rate. We refer to Table \ref{astrodetails} for details of final lens settings and related parameters. 

It is interesting to note that in practice the lens allows for focal length shorter than 4.5 mm (as announced on the enclosing, cf. Fig.~\ref{camandlens}). The lenses of camera 3 
at both SOAO and BOAO observatory stations have a focal length of around 3 mm. 

For reference the focal length of a lens can be determined from the physical size of the CCD and field of view (from astrometry). If $W$ denotes the horzontal width of the field of view in degrees and $h$ is the horizontal size of the CCD, then 

\begin{equation}
\frac{W}{2} = \tan^{-1}\bigg( \frac{2h}{f}\bigg),
\end{equation}
\noindent
where $f$ is the lens focal length. For the Watec model in our setup $h=640 \times 8.4~\mu$m and the horizontal field of view was determined from astrometry (see forthcoming section 
on UFOAnalyzerV2) and can be looked up from Table \ref{astrodetails}. We note that the spatial resolution ranges from 4'/pixel to 7.7'/pixel. For comparison the resolution at the CAMS \citep{CAMS1} network is 2.8'/pixel. This is due to a smaller field of view of each camera at the CAMS network.

For identical sensor size it is expected that relatively larger field of view will yield a lower sensitivity (and hence a lower meteor detection rate) compared to a smaller field of view lens setting. Since this project to some extent relied on the successful detection of one or (preferrably) more double-station meteor events we opted for several larger field of views settings having in mind that the detection probability would increase with an increased sky coverage. A more professional but budget-intensive approach would have been to chose several small (partially overlapping) field of view cameras. This approach was follwed by the CAMS project \citep{CAMS1} in an optimized approach. However, an increase in the number of cameras at SOAO and BOAO would have required more time for the installation and a more generous budget. The described system in this paper represents a proto-type / feasibility study.

Each lens was then focused on the background stars during dark clear-sky conditions. Focusing is important for correct astrometric plate calculations. We experimented with the back-focus option available to this camera model, but could not find any differences. The final focus setting of each lens was fixed mechanically via fixing screws. 
An advantageous technical feature of the lens is the auto-iris function powered via a connector plugged into the Watec camera (cf. Fig.~\ref{camandlens}). Once the camera switches off the lens iris will automatically close and thereby protect the camera CCD from direct sunlight. The iris-level adjustment (manually adjustable by camera setting) was set to fully open the lens-iris diaphragm to ensure widest aperture for maximum light collection.

\subsection{Digital video-capture and peripheral equipment}

\begin{figure}
\centering
{\includegraphics[width=0.35\textwidth]{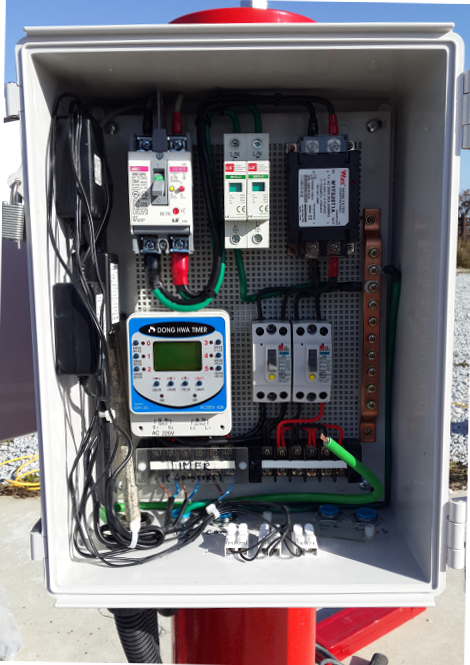}}
\caption{Detailed view of the power supply (main switch) box with automatic operation (Dong Hwa) timer (at SOAO). Black boxes on the top-left side are the three 12V power supply transformers for the Watec cameras. \emph{See online version for colors}.}
\label{powerbox}
\end{figure}

The video and power supply units were securely enclosed in two separate rain-proof hard-plastic enclosures (cf. Fig.~\ref{powerbox}) attached to the metallic vertical cylinder-shaped base-mount of each camera station. 

At each site the video-signal from each camera was transmitted via a single BNC-to-BNC 
(type 5C-HFBT, 75$\Omega$) coaxial video cable. The estimated camera-PC distance was 44m and 53m at SOAO and BOAO, respectively. The coaxial cable wires are running partially underground. Special hard-plastic tubes were used for further damage-protection of the coaxial cable. Each station was grounded to protect the equipment against damage due to 
lightening strike. However, high-voltage current created in the coaxial cables due to nearby lightening strikes are not safe-guarded by the grounding. The power supply was 
controlled by a professional programmable timer to automatically switch on/off the 12V power supply of the cameras. Initially, we installed surge-protection devices (SPD) at each end of the coaxial cable to protect (camera and PC) against damage due to the event of a lightening-strike. But abandoned this safety feature due to significant signal loss. 

During the initial phase of installation we experimented with various signal-amplification equipment mainly to mitigate possible signal-loss due to a long wiring distance. We found 
that amplification did not improve significantly the image quality even when tested under extreme cable lengths of 100 meter.

For the analog to digital (AD) video conversion we used three types of converters (also known as framegrabbers): a) EzCapture\footnote{http://www.ezcap.tv} 116 and currently in use at the BOAO station, b) Hauppauge\footnote{http://www.hauppauge.com} Impact-VCBe PCI card (BNC to S-video connector required) which were destroyed in 2015 during a lightening strike and c) Pinnacle Dazzle\footnote{http://www.pinnacle.com} DVC100 currently in use at the SOAO site. All AD video converters produce a final image size 640 x 480 pixels at a framerate of about 30 frames per second and therefore do not introduce any important changes to address. Any capture device that maintains these technical specifications can be used without any effects or alteration of the camera-specific astrometric plate solution (see forthcoming chapter on UFOAnalyzerV2). We find that custom image size settings are not possible for any of these capture devices. However, high-end video-capture PCI or PCIe based cards are likely to offer image size selection which could improve the astrometric resolution accuracy. We find that the AD converters require their own device drivers for secure parallel operations of multiple data income streams. For this reason we installed a USB-PCI card in the SOAO detection PC with its own USB device driver (recognised automatically by Win 7 OS). However, although many AD capture devices with individual device drivers can be installed in a single PC (many PCI/PCI-e slots available), the computing power will set a limit to the correct operation of each camera. In case of system-overload frame-drops might occur in the recorded event video. An issue that should be monitored and addressed for correct trajectory determination.

\begin{figure*}
\vbox{
\centerline{
\includegraphics[scale=0.25]{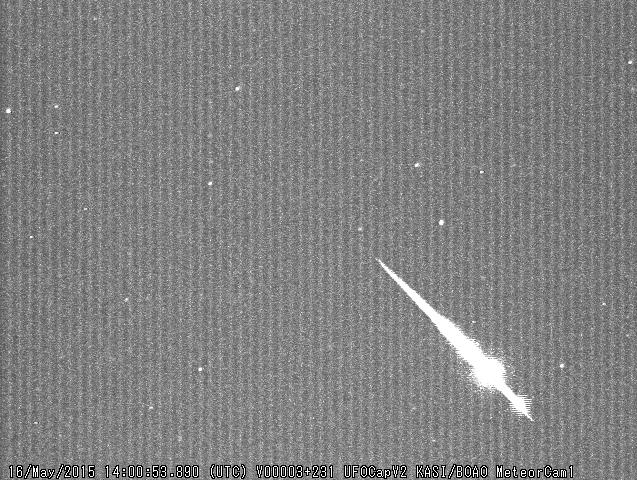}
\vspace{0.5mm}
\includegraphics[scale=0.25]{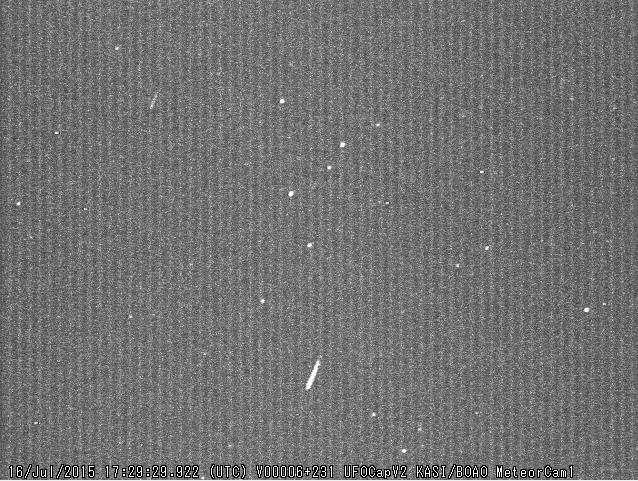}
\vspace{0.5mm}
\includegraphics[scale=0.25]{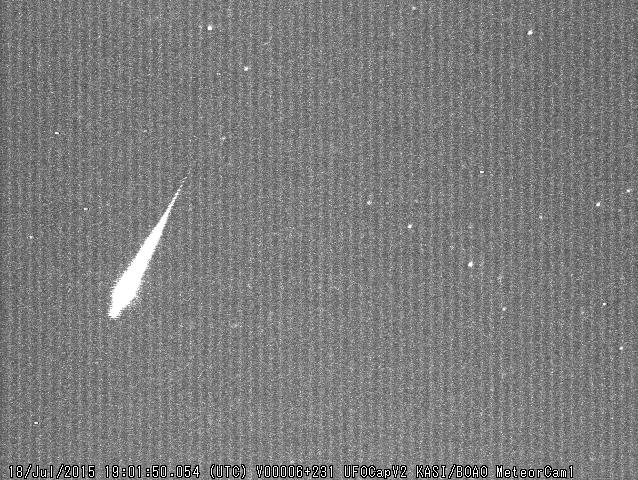}
}}
\vbox{
\centerline{
\includegraphics[scale=0.25]{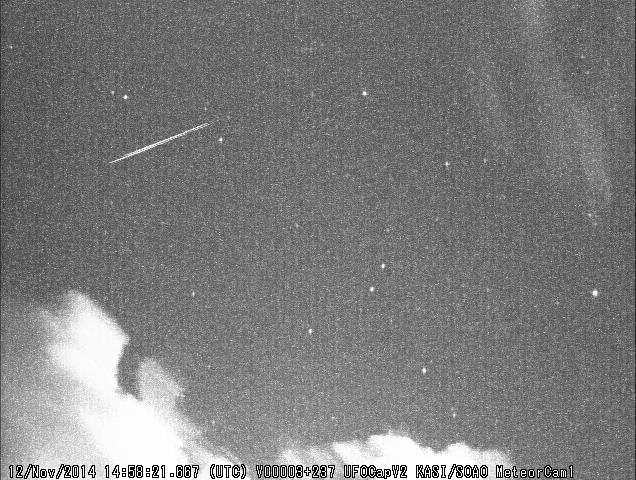}
\includegraphics[scale=0.25]{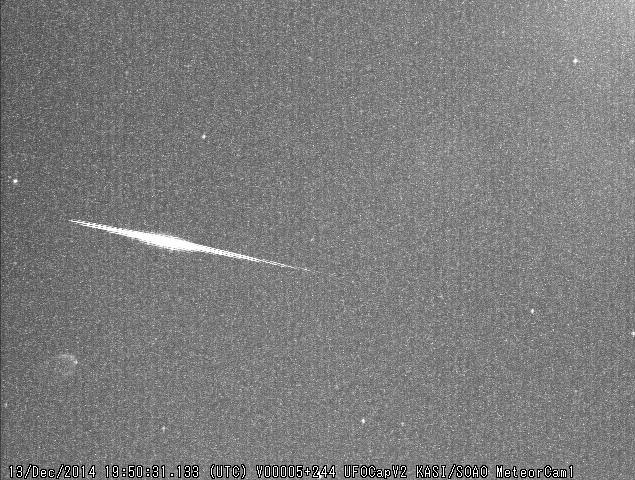}
\includegraphics[scale=0.25]{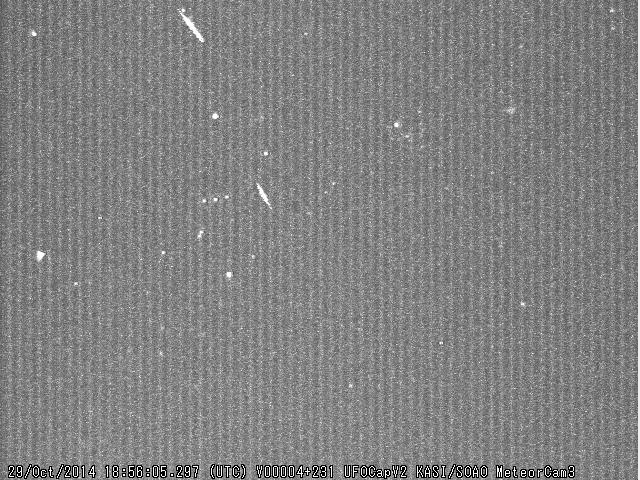}
}}
\caption{Various detections of faint and bright meteors at SOAO and BOAO. Top panels are for BOAO (Cam1, Cam1, Cam1). Bottom panels for SOAO (Cam1, Cam1, Cam3). For the detection at BOAO camera 1 (top-middle panel) the constellation of Cassiopaia is clearly seen. The bottom-right panel shows two consecutive meteor detections with 1 second apart from each other. Due to the large field of view the constellation of Orion is clearly visible. Both meteors share a common direction of origin and were classified as members of the Orion meteor shower. Time stamps (all UTC) from top (left to right): 2015-05-16@14:00:53.89, 2015-07-16@17:29:29.92, 2015-07-16@19:01:50.05, 2014-11-12@14:58:21.67, 2014-12-13@19:50:31.13, 2014-10-29@18:56:05.29.}
\label{variousdetections}
\end{figure*}

\begin{figure}
\centering
{\includegraphics[width=0.40\textwidth]{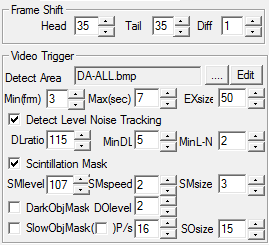}}
\caption{GUI settings for the input tab within the UAV2 software. The most important settings to our experience is the scintillation mask, noise level tracking. Dark and slow object masks were not enabled. We note that these settings are not optimized for detecting satellite tracking motion. Also no systematic study was undertaken to find optimized values. \emph{See online version for colors}.}
\label{UCV2GUI}
\end{figure}

The computing architecture is an Intel Core-i7 (3.4 GHz) at SOAO and a Intel Core2 Quad CPU (2.67 GHz) at BOAO. Initially, both PC's were identical to the BOAO architecture. However, a lightening strike in 2015 damaged the mainboard of the SOAO PC permanently and was subsequently substituted with a replacement PC. All data processing and video-analysing computers were 
kindly sponsored by KASI.

Three video feeds (from three cameras) are processed on each PC. While the BOAO PC works flawless since 2014, at times, we experience technical problems with the replacement PC at SOAO. At times an unexpected blue-screen of death (BSD) error which causes the PC to shut-down and reboot. We suspect that the error is connected with a USB bus, but due to time limitation, we were not fully able to identify the root-cause of this failure. Another cause for this error could be PC cabinet over-heat (\emph{SonotaCo private communicatioin}). This still has to be tested starting from the fall 2017 season. In the case of a BSD power cycle all ongoing observations are terminated instantly and can only be resumed by a manual restart of the meteor detection software requiring re-attachment of each video capture-device driver. Each PC is running a Real VNC\footnote{http://www.realvnc.com/} server for purpose of remote login for maintanance work. To ensure highest CPU performance we enabled a custom energy and work-load saving settings. Automatic updates of the operating system are only performed manually at the request of the system administrator. Further more disk intensive maintanance work is automatically scheduled to start in day-time hours and hard-drive spin-down is permanently disabled. For data backup purpose we installed the versatile GoodSync\footnote{http://goodsync.com} FTP-based data transport and synchronisation program. Nightly data were then 
transferred during daytime to the main meteor PC located at KASI/Daejeon. Data and other information can be retrieved at http://meteor.kasi.re.kr.

\section{Data acquisition and processing \label{sec:dataproc}}

For the detection of atmospheric meteor events we made use of Sonotaco's UFO\footnote{http://sonotaco.com} software package. It was earlier used to compile a meteor shower catalog based on observations made in 2007-2008 \citep{sonotaco2009, kanamori2009}.

Three branches are available in this software suite each with its own functionality and purpose: i) \emph{UFOCaptureV2}, ii) \emph{UFOAnalyzerV2} and iii) \emph{UFOOrbitV2}. Practical experience in working with this software were made previously by the lead author at the Armagh Observatory, UK. We find this software to be very stable and relatively userfriendly. The software operation is straighforward and relatively easy to use even for first-time beginners and was judged to be suitable to ensure a success of this proto-type pilot study. The advantage is that it can be setup without much trouble within thirty minutes (under guidance) and the user licence fee is reasonable. Korean amateur astronomers living in rural areas 
are likely to catch their own video-meteor within a few days after setting up there own camera system.

In the Republic of Korea the software has not been used within the korean meteor detection community and we think its beneficial to give a more deeper description which could serve as an introduction to its use for future first-time users. 

However, we refer to the latest online software documentation which is updated on a frequent basis and contains a Q\&A section with answers on most relevant questions. For readers who are interested in detailed settings are invited to contact the corresponding author for such information. Screenshots of windows of particular software parameter settings can be provided and guidance towards an optimized independent system offered.

\subsection{Meteor capture with ``UFOCaptureV2''}

The main detection software is embedded within the UFOCaptureV2 (\citet{UAV2}, V2.24 and named UCV2 hereafter) program. At first glimpse the graphical user interface (GUI) looks overwhelming. However, the main functionality condenses down to a few settings and the documentation (available in English and Japanese) is well explained and structured. In addition, the software author is maintaining an active discussion forum where questions can be posted. The program runs exclusively on windows-based PC's starting from Microsoft Windows 2000, XP, Vista and upwards. The meteor PC's at BOAO and SOAO are currently running Windows 7 (SP2).

UCV2 is a time-shift-motion-detect-video-recording software. UCV2 is compatible with a variety of video capture-devices interfacing via USB, firewire (IEEE 1394) or analog video (converted to digital). Most users make use of feeding the analog video-signal (in the present case the Watec camera) into a USB enabled AD converter. The main requirement for the capturing device to be recognized by UCV2 is DirectX compatibility. 

Both brightness, temporal and area size detection video-recording triggers can be utilised for motion detection. As a proof-of-concept Fig.~\ref{variousdetections} shows examples of 
the recording of various meteors obtained with the SOAO/BOAO cameras. 

In Fig.~\ref{UCV2GUI} we show the input sheet of the software GUI. These features renders UCV2 to be very versatile and finds application from bird tracking to the detection of meteors and satellites (with suitable optical equipment and software settings). For meteor detection a pixel-to-pixel brightness change is enabled due to its discerning changing nature in brightness across the CCD chip. 

\begin{figure}
\centerline{\frame{\includegraphics[scale=0.37]{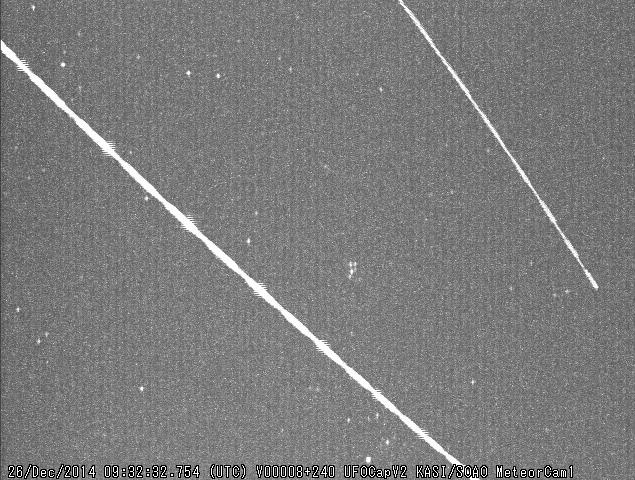}}}
\vspace{1mm}
\centerline{\frame{\includegraphics[scale=0.37]{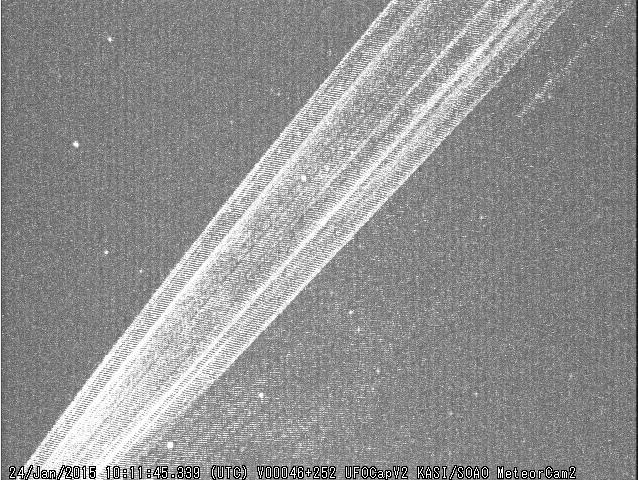}}}
\vspace{1mm}
\centerline{\frame{\includegraphics[scale=0.37]{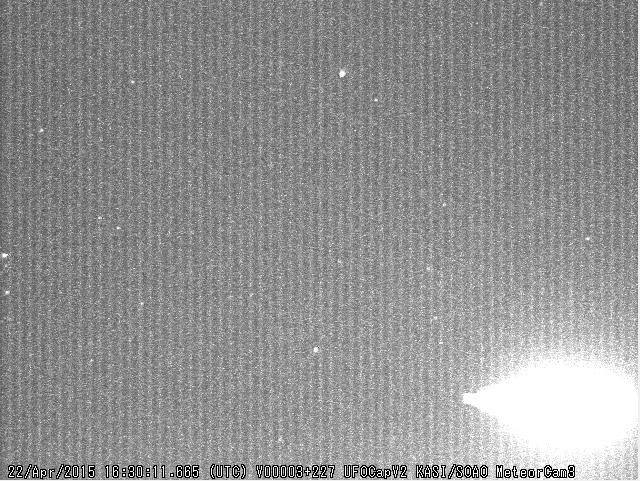}}}
\caption{Non-meteor detections of moving celestial phenomena. \emph{Top panel:} airplane tracks (constellation of Pleiades is visible). \emph{Middle panel:} movement of powerful laster pointer (public outreach at SOAO). In addition a possible satellite is tracked against the background stars, but this is only visible in the video. 
\emph{Bottom panel:} Unusual bright fireballs and/or iridium flashs. Most false-positive detections are due to airplanes.}
\label{nonmeteor}
\end{figure}

Several trigger options exist to eliminate false-positive detections such as moving airplanes, satellites or birds/insects. For their elimination dark/slow object masks can be enabled. For stationary but temporarily moving objects (trees and/or bushes) that are present in the field of view the manual free-hand area mask is suitable to exclude certain areas on the CCD from the triggering feature. To define free-hand area mask a special purpose editor will assist to define such areas. In our setup all cameras at SOAO/BOAO have an unobscured field of view. Atmospheric effects that could cause pixel-to-pixel variations are handled by a scintillation mask. Fixed background stars 
that exhibits some degree of scintillation are then detected by the software and tracked and consequently disregarded as a detection trigger. Triggering options can also be based on the duration of the triggering event. Short-duration cosmics events are omnipresent causing star-like objects or streaks to appear on the CCD. Another short-duration trigger could be interference caused by bad or loose connections in the signal wiring. For such events a minimum frames setting can be enabled and used to exclude events that are longer in duration in units of frames (approximately 30 
frames = 1 second). These settings added together reduce the amount of false-positive detections substantially. However, at times such events (airplanes, etc.) are detected and their subsequent manual removal necessary. In Fig.~\ref{nonmeteor} we show the recordings of a few examples of false-positive triggers. We speculate/conjecture about the detection of satellite tracks. However, they are not actively detected due to their poor photometric/kinematic properties (faint and slow). When the cameras detect satellite tracks its often by chance when the recording was triggered by an airplane or an actual meteor.

\begin{figure*}[!ht]
\hbox{
\centerline{
\includegraphics[scale=0.375]{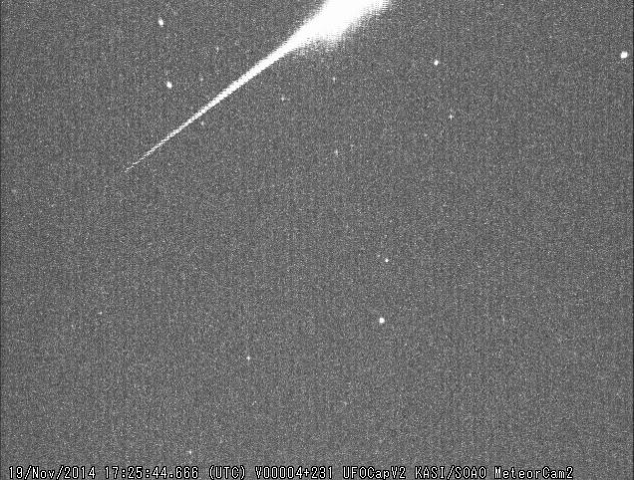} 
\hspace{0.5mm}
\includegraphics[scale=0.375]{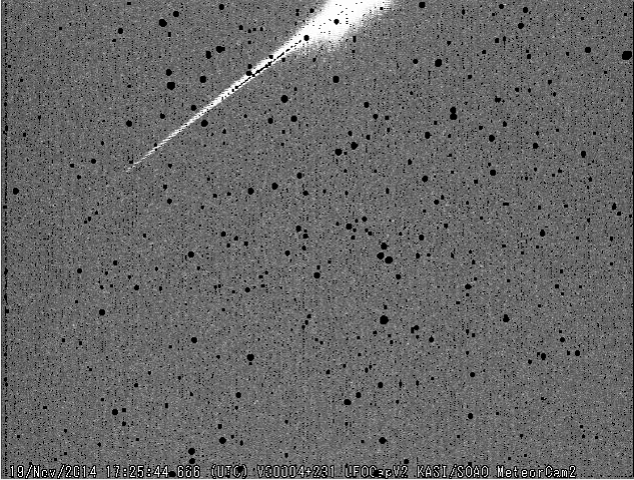} 
}}
\vspace{2.5mm}
\hbox{
\centerline{
\includegraphics[scale=0.375]{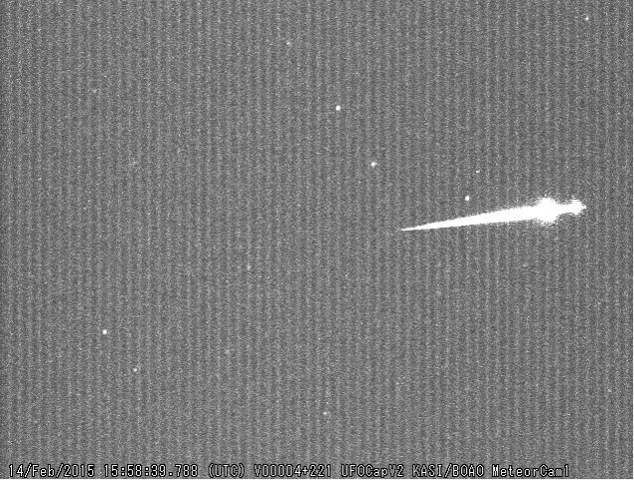} 
\hspace{0.5mm}
\includegraphics[scale=0.375]{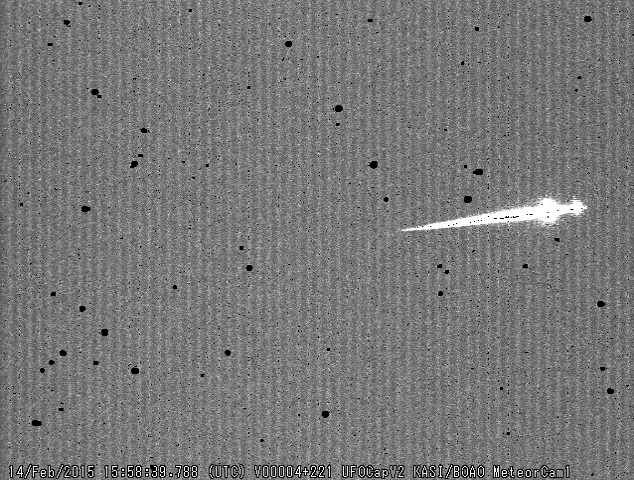} 
}}
\caption{Examples of the recording of two different bright fireballs at SOAO (top, camera 2) and BOAO (bottom, camera 1) observatories. Masked (black dots) background stars are shown in the two right-hand panels. Hot pixel defects are also masked and the corresponding MS-link (see text) need to be removed during the astrometric calibration process. Background star-mask positioned on top of the meteor event will also be included in the astrometry calculations. We note that some background stars are detected by the software and not recognised visually. For BOAO camera 1 vertical interlace stripes are seen.}
\label{twodifffb}
\end{figure*}

The video recording on disk is done via utilisation of internal memory buffering. Depending on settings the buffer contains video information of the past. The buffer stores information in units of frames. For example a heading (trailing) buffer of 60 (30) frames would start to record approximately 2 (1) seconds before (after) the trigger event.

Finally, UCV2 offers at wide array of settings for housekeeping purposes and automatic operation. 
Details on capturing device, camera, lens model and site location are conveniently stored in the profile settings menu. An operation timer is available 
to automatically start and stop the UCV2 software taking into account Sun rise/set times. Clips of recorded videos can be viewed in specified date intervals or deleted alltogether or one-by-one if necessary. Scintillation or user-defined masked can be superimposed on the video. An automatic restart interval option is available as well for systems that are unstable. This has the advantage that the UCV2 software does restart and reinitiate all necessary settings after a power-cycle.

\subsection{Astrometric calibration with ``UFOAnalyzerV2''}

UFOAnalyzerV2 (\citet{UAV2}, V2.07, UAV2 hereafter) is concerned with the analysis of data generated by the UCV2 detection software. In particular UCV2 generates a *.avi (event movie), a *.xml (contains observing station and equipment information) and a *.bmp (star-mask) file of which all are necessary for astrometric post-processing with UAV2. For a given hardware setup (camera + lens + capture device\footnote{the capture device might change the number of pixels}) UAV2 creates a unique hardware profile database. In its very essence UAV2 performs a transformation of image coordinates to standard equatorial coordinates by determining an astrometric plate solution (field of view center, scaling and optical distortion parameters) for a given meteor recording. We refer the reader to the excellent monograph by \cite{montenbruck} for details on astrometry and orbit determination. For each camera we have carried out an astrometric calibration and details are shown in Table \ref{astrodetails}.

The UAV2 software makes use of the scintillation masks detected by UCV2 and compares their positions with a star catalog    
(Sky2000\footnote{http://tdc-www.harvard.edu/catalogs/sky2k.html}) containing around 300,000 stars brighter than 8th magnitude. The star catalog is superimposed on the recorded (layered) meteor image. The distance of so-called mask-star (MS) links (displayed graphically) are then attempted to be minimized over two steps. The first step is relying on initial parameter settings provided by the user and involves educated trial and error guesses of geographic location/orientation as well as hardware parameters. Special attention should be paid on the horizontal field of view, the azimuth, altitude and rotation angle parameter of the camera (cf. Table \ref{astrodetails}). The second step, resulting in a more refined astrometric solution, is done by software via iterative least-squares minimisation on the residuals of the MS-on-sky-distance. Higher-order effects such as lens aberration constants ($k_2, k_3, k_4$, cf. Table \ref{astrodetails}) can be included during this step to refine the astrometric solution (especially for stars located in the near-edge of the field of view). In Fig.~\ref{twodifffb} we show two meteor events and their corresponding masked background stars. 

The first step is essential in order to provide reasonably close initial guesses for the minimization algorithm to converge towards a more accurate solution and is essential in the automation process of finding an astrometric solution. The accuracy of the astrometry is measured in pixels (converted to an angle) between the brightest point of the CCD and the reference star's coordinates. This accuracy can be improved by eliminating a subset of MS-links with a standard deviation larger than the average standard deviation of all established links. CCD defects like hot pixels or cosmics events could mistakenly be identified as catalog stars and should be avoided. To build intuition on the use of UAV2 we recommend experimention by the deletion of MS-links one by one and record the resulting accuracy. For obvious reasons not all stars (possibly defects present) can be used. On the other hand including only a hand-full of stars is not advisable either. MS-links should be chosen across the whole field of view of the CCD in order to avoid finding local minima in the minimisation process. As a rule-of-thumb around 50 MS-links should be retained for a final accurate astrometric plate solution. In this study, we routinely achived an average 
astrometric error of less than 0.5 pixels (cf. Table \ref{astrodetails}).

One thing is worth to remember. Nearby bright stars might have a proper motion and will change their position on the sky over time. The applied star catalog therefore should be updated on a regular basis to account for changes of astrometric positions. However, if the proper motion change is smaller than the average astrometric residual between measured/observed stars and the corresponding catalog object, then proper motion may be neglected.

\begin{figure}
\centerline{\includegraphics[scale=0.215]{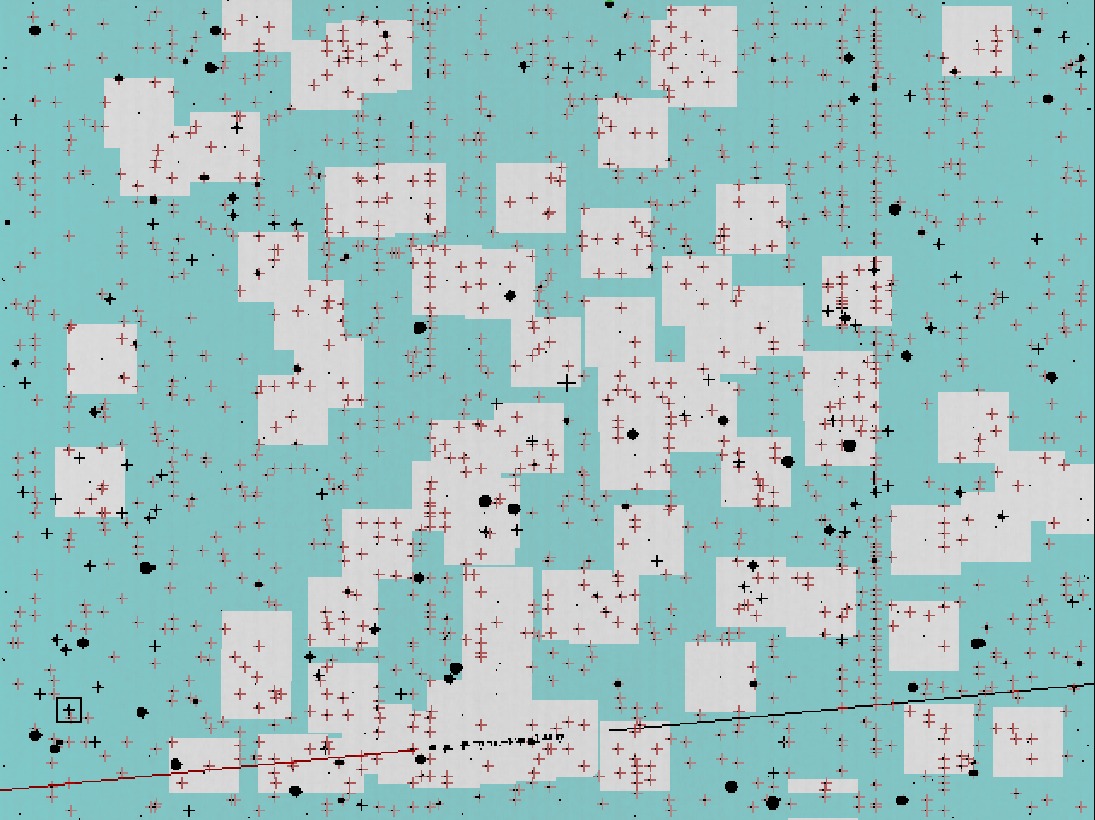}}
\caption{Graphical representation (inverted colors) of the final astrometric solution for event 
2014-11-18 18:59:29 (UTC) at BOAO camera 2. Red crosses mark superimposed masks. Black crosses mark position of 
background stars. Some stars have a broad PSF and therefore hide the cross. The meteor trail 
is along the straight line. The box in the lower left corner
is the mask-star pair that has the largest position error (0.67 pixels). Also the CCD seems 
to have a bad pixel column (vertical direction) on the right hand side. The large cross in 
the middel of the figure is the astrometric center of the field of view. See Table 
\ref{astrodetails} for astrometric data of all cameras. \emph{See online version for colors}.}
\label{astrometry}
\end{figure}

Once hardware and orientation specific parameters (field alignment) have been established for a given camera, it is now possible to measure meteor specific properties such as speed and position. The detection of photometric meteor trail centroids is carried automatically and the meteor trail found from linear regression. In Fig.~\ref{astrometry} we show an example of a final astrometric plate solution and we refer to the figure caption for details. To determine the position (altitude and azimuth or right-ascension and declination) of the meteor the beginning and end points of the meteor trail are requested and found graphically by the user. The analyze section of UAV2 then allows to measure the direction, speed, linearity and magnitude 
of the meteor. In case of a single-station meteor detection the distance can be calculated from the observed altitude and an assumed height. For double-station meteors the height assumption is eliminated.

Additional features are available from within UAV2. For several meteor detections UAV2 allows batch job processing on all events without manual user interfacing of individual detections. This feature is time saving and allows the processing of several data sets in seriel (for example for data accumulated over a complete season). The ``class'' function allows the classification of a meteor to a possible meteor stream or as a sporadic event. The astrometric solution for each event in batch mode is reliable since each camera was ``calibrated'' and the initial guess of astrometric parameters is very close to the final parameters.

The ``plot'' function allows the drawing of field of views projected on ground maps. This function is helpful in obtaining an idea of the percentage overlap of the field of view of two cameras forming part of a double-station system. Another useful function is the drawing of meteor start-to-end trails projected on the ground.

\subsection{Meteor orbit determination with ``UFOOrbitV2''}

UFOOrbit (\citet{UOV2}, V2.41, hereafter UOV2) is the final branch of the Sonotaco meteor data analysis package. It allows the computation and visualisation of the physical orbit of a meteor event. For obvious reasons single-event data from a double-station observing network is necessary as a minimum. The accuracy of orbit computation improves with the increase in the number of detection stations. Based on time-stamp information UOV2 has the ability to automatically detect double-station events. Therefore, the user is not explicitly asked to provide double-station data. However, from experience, we find that the time-stamp at times is different for some events by up to one second. Especially this was the case for detections which were made early in the project initation. We then decided to increase the synchronisation frequency of the time synchronisation software among other settings. We refer to the next section for more details on time synchronisation at the two stations which is an integral part in double-station meteor observation.

\begin{figure*}[!ht]
\centerline{\includegraphics[scale=0.50]{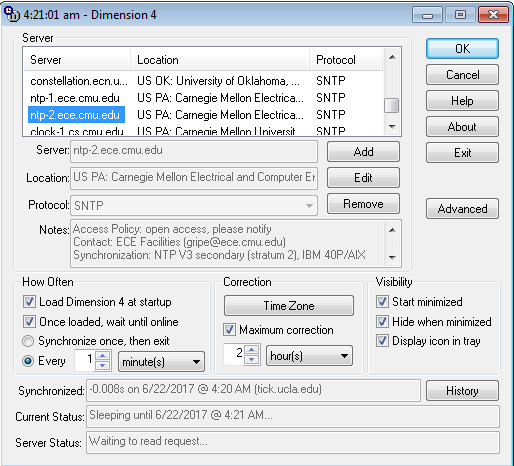}
            \includegraphics[scale=0.50]{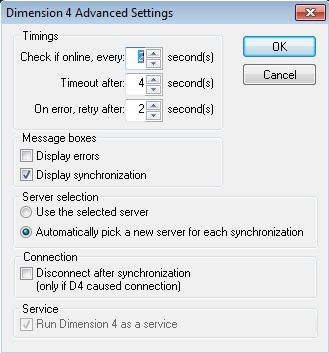}}
\caption{Graphical user interface (GUI) of Dimension 4 time synchronisation settings. \emph{Top panel:} Main GUI window. The software is started on each boot to ensure continuous synchronisation of the internal PC clock. Synchronisation is done at a 1 minute cadence. The timezone of the PC should be set to coordinated universal time (UTC). \emph{Bottom panel:} GUI of advanced settings option. The online checking cadence is set relatively high (2 seconds). We find that the most important setting is the automatic selection of a new time server. We found that some nearby time servers in Korea are offline for an extended time period. \emph{See electronic version for colors}.}
\label{sync}
\end{figure*}

The program calculates the full three-dimensional orbit and provides osculating Keplerian elements ($a, e, I, \omega, \Omega, M$) at the time of detection. In a multi-station network the meteor radiant point (important for cluster association determination) is calculated based on data from all (possible) pairs of stations. For a given meteor trail a 3D plane can be calculated originating at the station. The normal vector for each ``station'' plane can then be determined. The radiant is then determined from the cross-product of pairs of normal vectors. Increasing the number of station and by averaging over all pairs of plane intersections one obtaines an accurate estimate of the radiant point. With the help of an on-line database (provided by the International Meteor Organisation) the meteor is classified to a specific meteor stream (if known). For orbit computation a quality check filter can be invoked based on several parameters. This feature could be advantageous if accuracy of stream membership is required for more stringent classification. Based on all data the plotting section of UOV2 allows the visualisation of the radiant point of each detection on a Sanson map. The orbit plotting section allows the plotting of the orbit as well as displaying the results of a forward integration (specified by user input) in time under the general perturbation of all planets in the solar system. A slight short-coming of the UOV2 software is the calculation of a debris impact field on the ground accompanied with a recovery probability ellipse based on atmospheric conditions. This feature would enable a search for a possible meteor-fall within a relatively narrow geographic location.

\subsection{Local time synchronisation and event timing}

The UCV2 software uses the internal hardware clock of the video-capturing PC and superimposes the time-stamp of detection at the bottom of each recorded video. In the recorded video each frame is labelled with a time-stamp marking the beginning of event and enables the timing of any point in the meteor trail. The UCV2 capture software allows the setting of size and position as well as the precision of the time stamp. 

We chose to display time stamps with a 0.001 seconds precision although a realistic timing error is on the 0.5-second level. The time stamp string is later recognized by UAV2 for further data processing and is important in the determination of a three-dimensional orbit for a double-station detection. However, the time string can be extracted by other software means \citep{atreyaIDL} in case video meteor data is analyzed with the use of alternative software. 

For time-synchronisation we chose to make use of the the Dimension4\footnote{http://www.thinkman.com/dimension4/} freeware software. For proper functioning the PC's time setting has to follow the UTC time standard and the time zone has to be changed accordingly. In UAV2 it is important to set the time-zone in accordance to the time-zone used when observations are recorded. Additionaly the MS Windows 7 internal time synchronisation setting has to be disabled. In the present settings we chose to synchronise every 1 minute. Further we chose an automatic server selection at each synchronisation. In Fig.~\ref{sync} we show the GUI of the Dimension4 main settings as well as advanced settings option. 

At first thought one might think that the best choice would be to select a time-server closest to the observing station. However, this limits the synchronisation to only one server. In case of server failure time synchronisation might become off-sync. The software runs flawless and a logging of past synchronisations is available in form of a ASCII logging file. 
Other time synchronisation software exist \cite{koschny2014}. In the future we plan to install on-site GPS based timing equipment for enhanced timing precision.

Some words on the timing of the event is at place. The event timing is of great importance for the final trajectory calculations including the determination of the meteors velocity vector. In UCV2 the time stamp is superimposed on the recording AVI movie and is initiated ahead of the actual meteor event determined by a user setting (head/trail parameters in Fig.~\ref{UCV2GUI}). Some issues might arise that one need to be aware of. Sometimes, by coincidence, the meteor event is preceded by another triggering event (a plane or cosmic ray hit or electrical noise or even a different meteor event - see Fig.~\ref{variousdetections}). In this case the timestamp assigned by UAV2 might be erronouse and the event (in case of double-detection) will never appear as in the directory listing in UOV2. 

\section{First results \label{sec:results}}

In this section we will present a few quantitative results of mostly statistical nature that have emerged from operating the cameras for the past 
2.5 years. 

\begin{figure*}[!ht]
\hbox{
\centerline{
\includegraphics[scale=0.70]{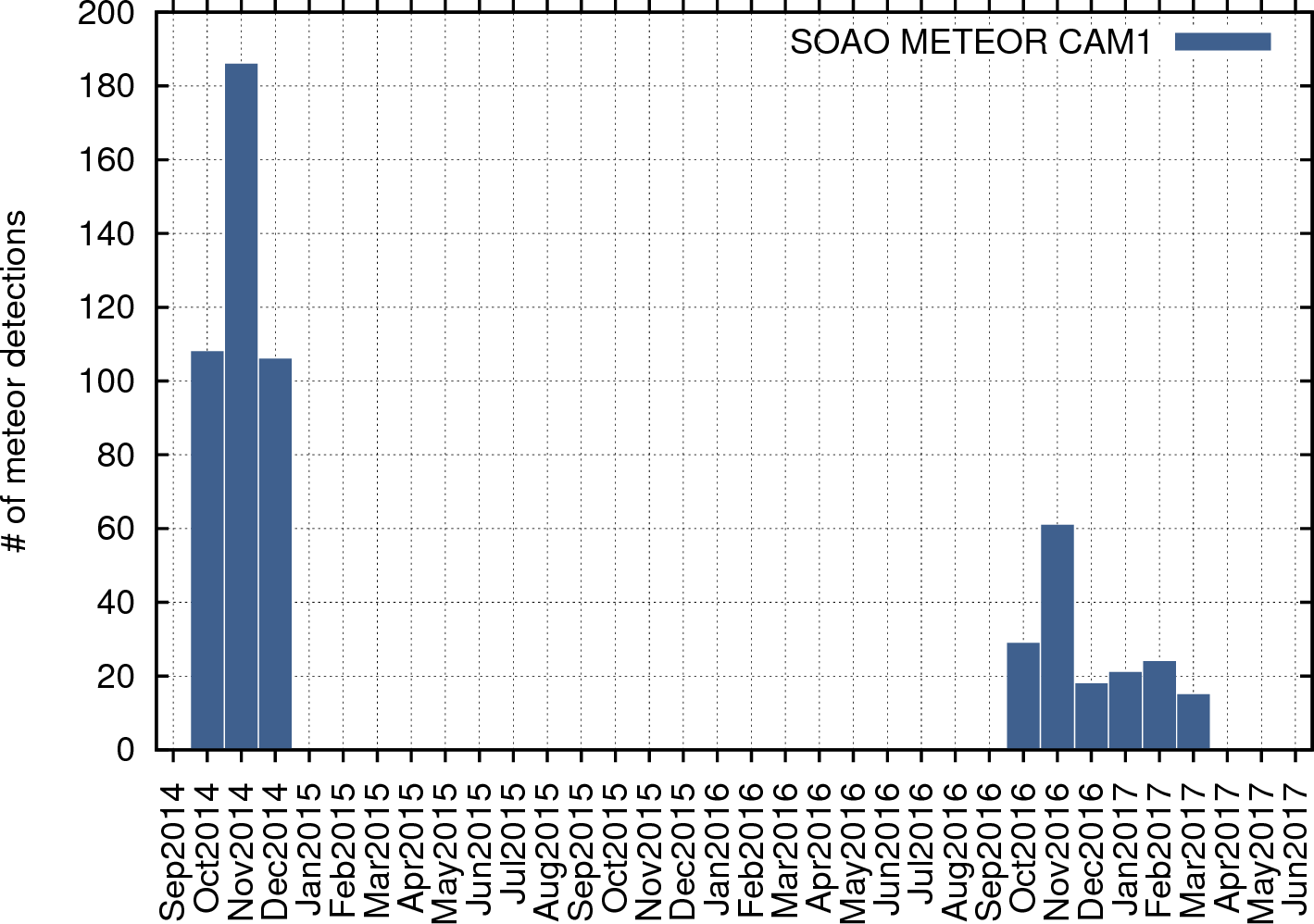}
\hspace{0.5mm}
\includegraphics[scale=0.70]{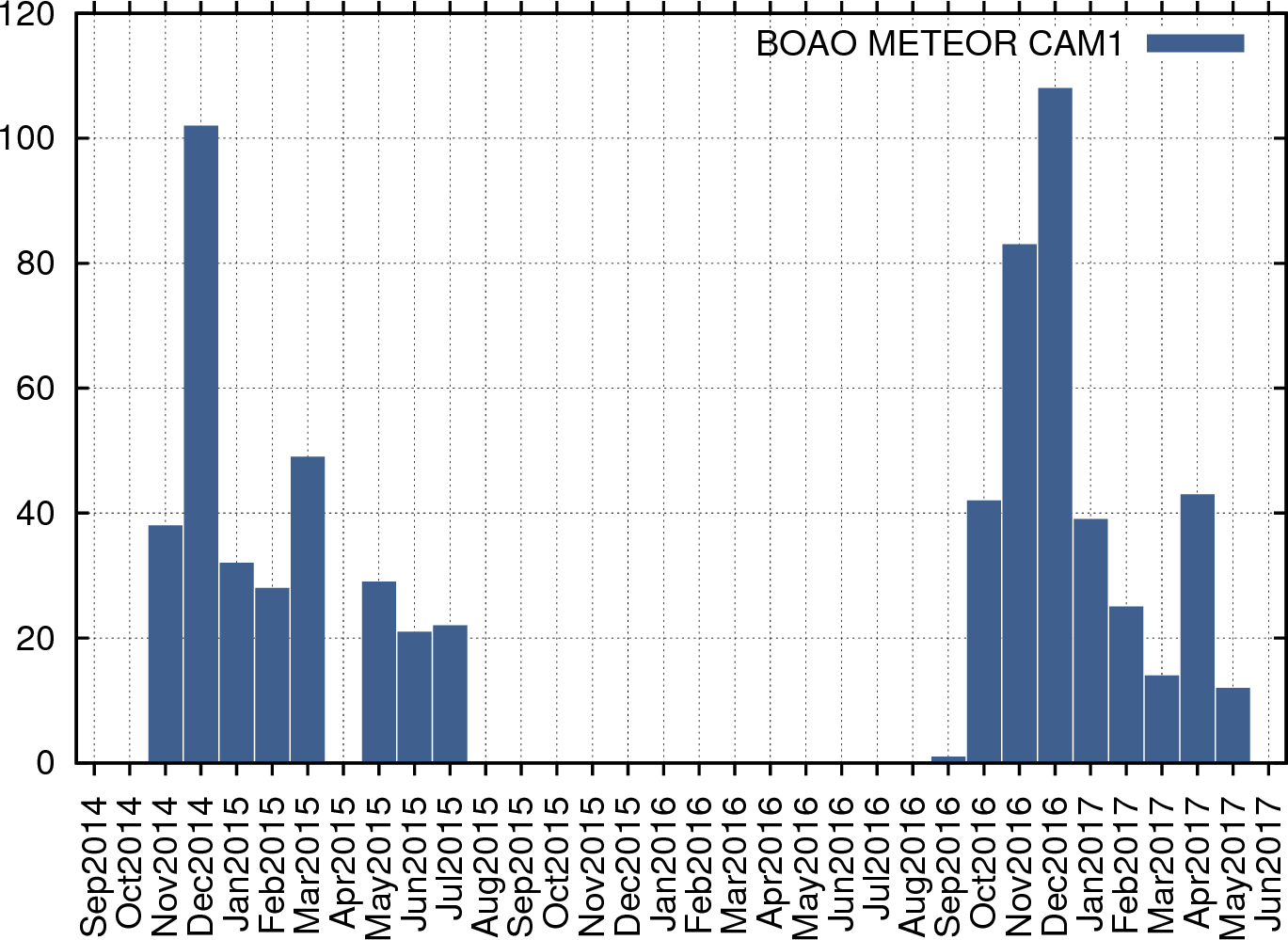}
}}
\vspace{2.5mm}
\hbox{
\centerline{
\includegraphics[scale=0.70]{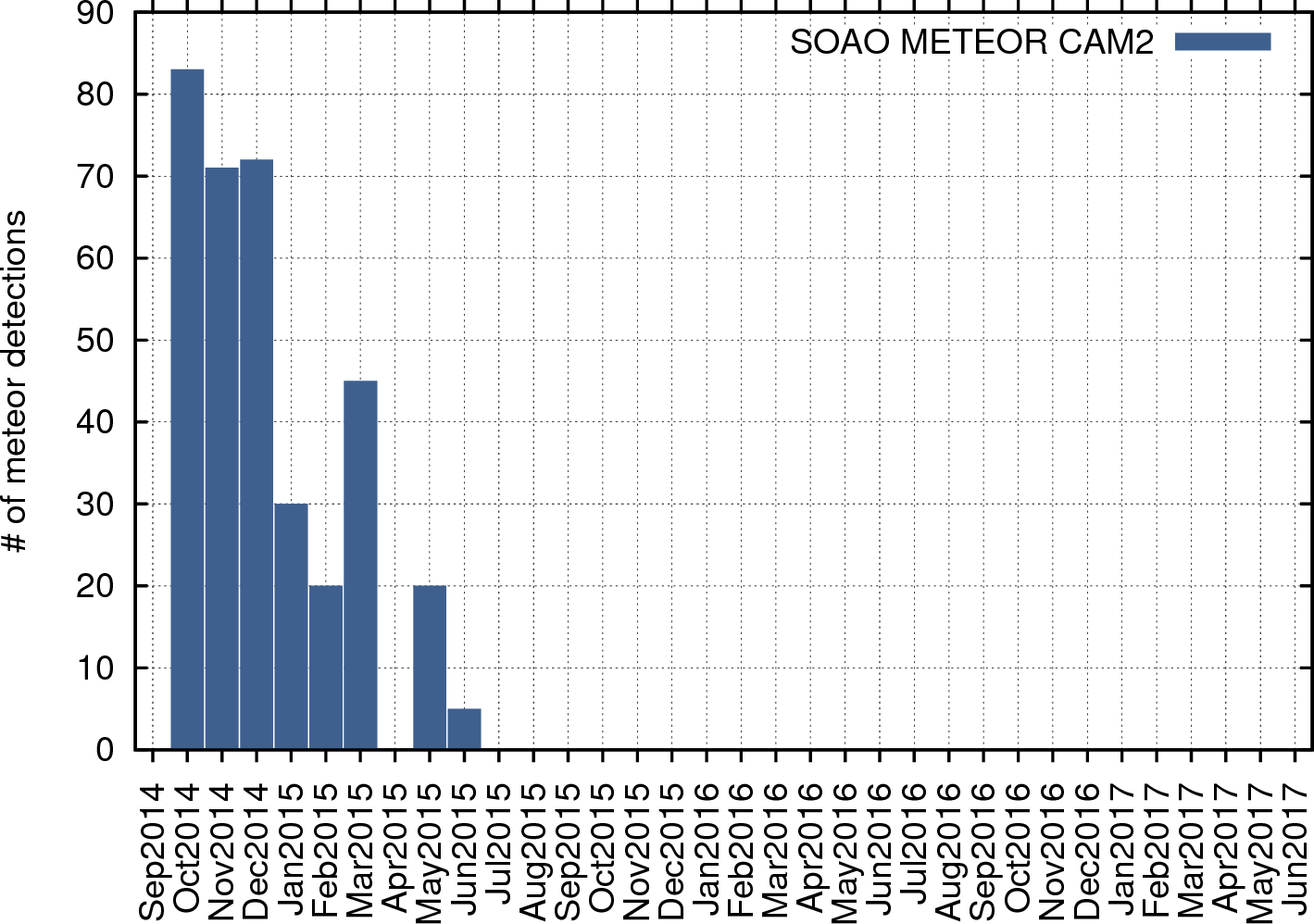}
\hspace{0.5mm}
\includegraphics[scale=0.70]{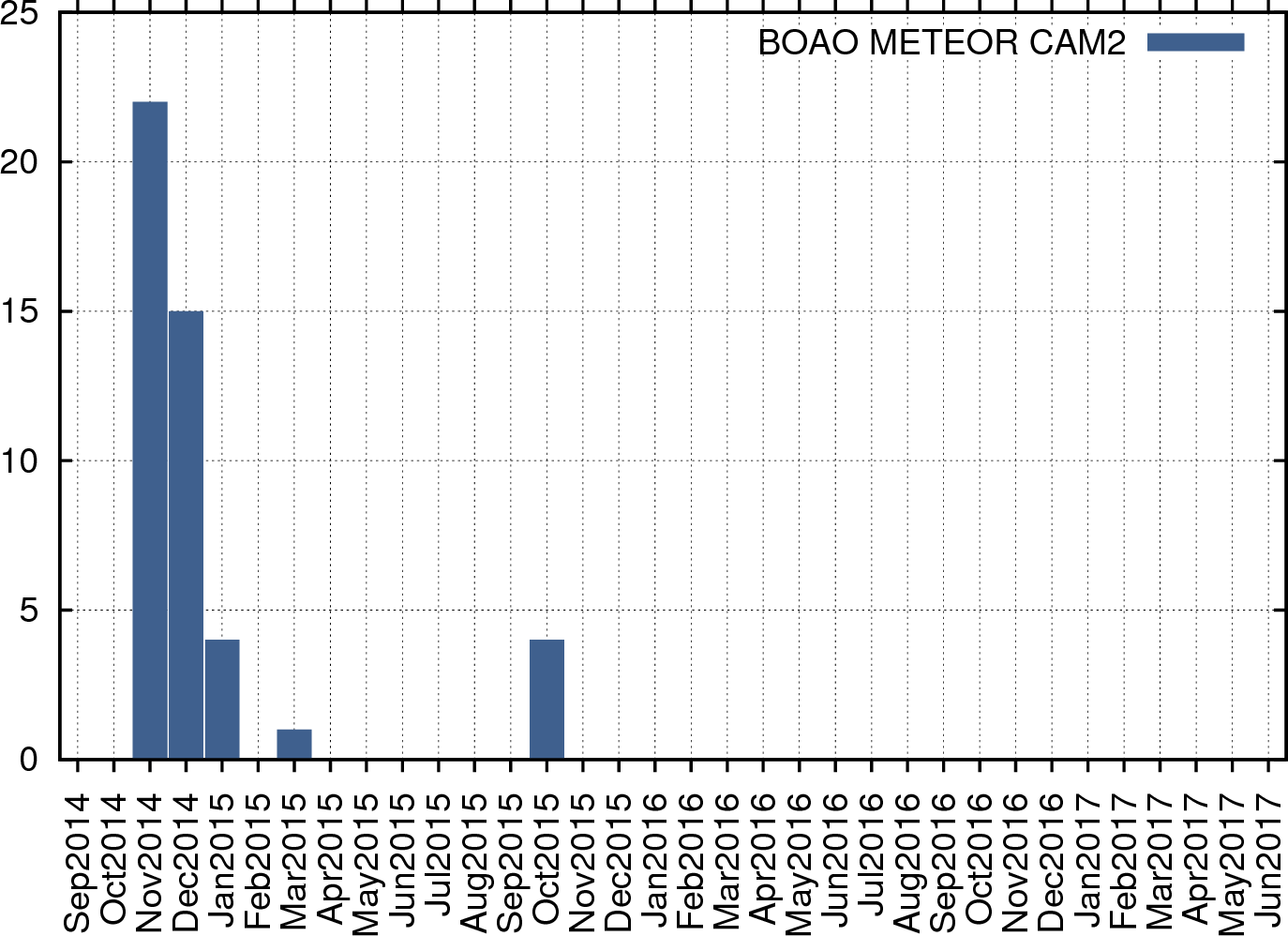}
}}
\vspace{2.5mm}
\hbox{
\centerline{
\includegraphics[scale=0.70]{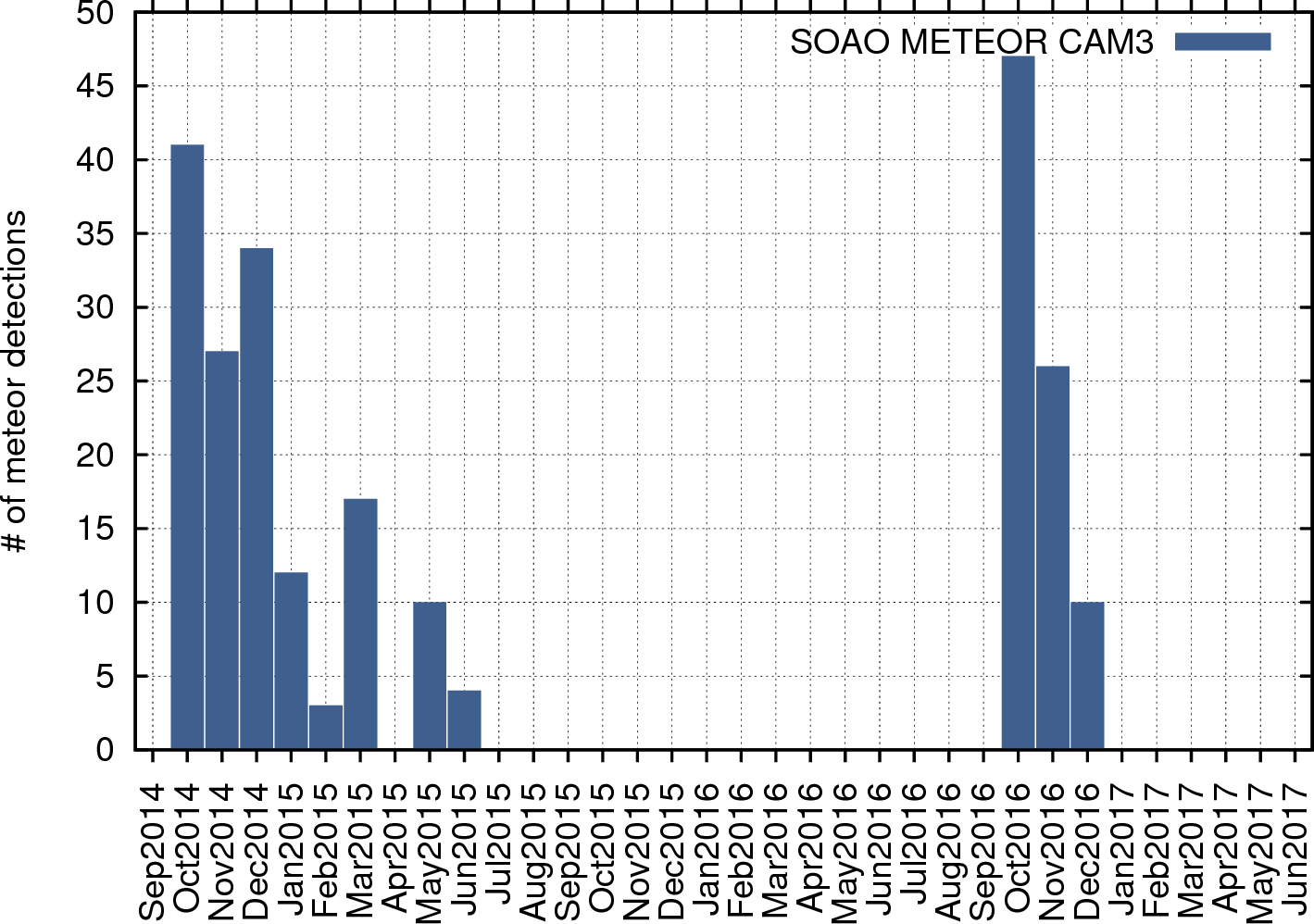}
\hspace{0.5mm}
\includegraphics[scale=0.70]{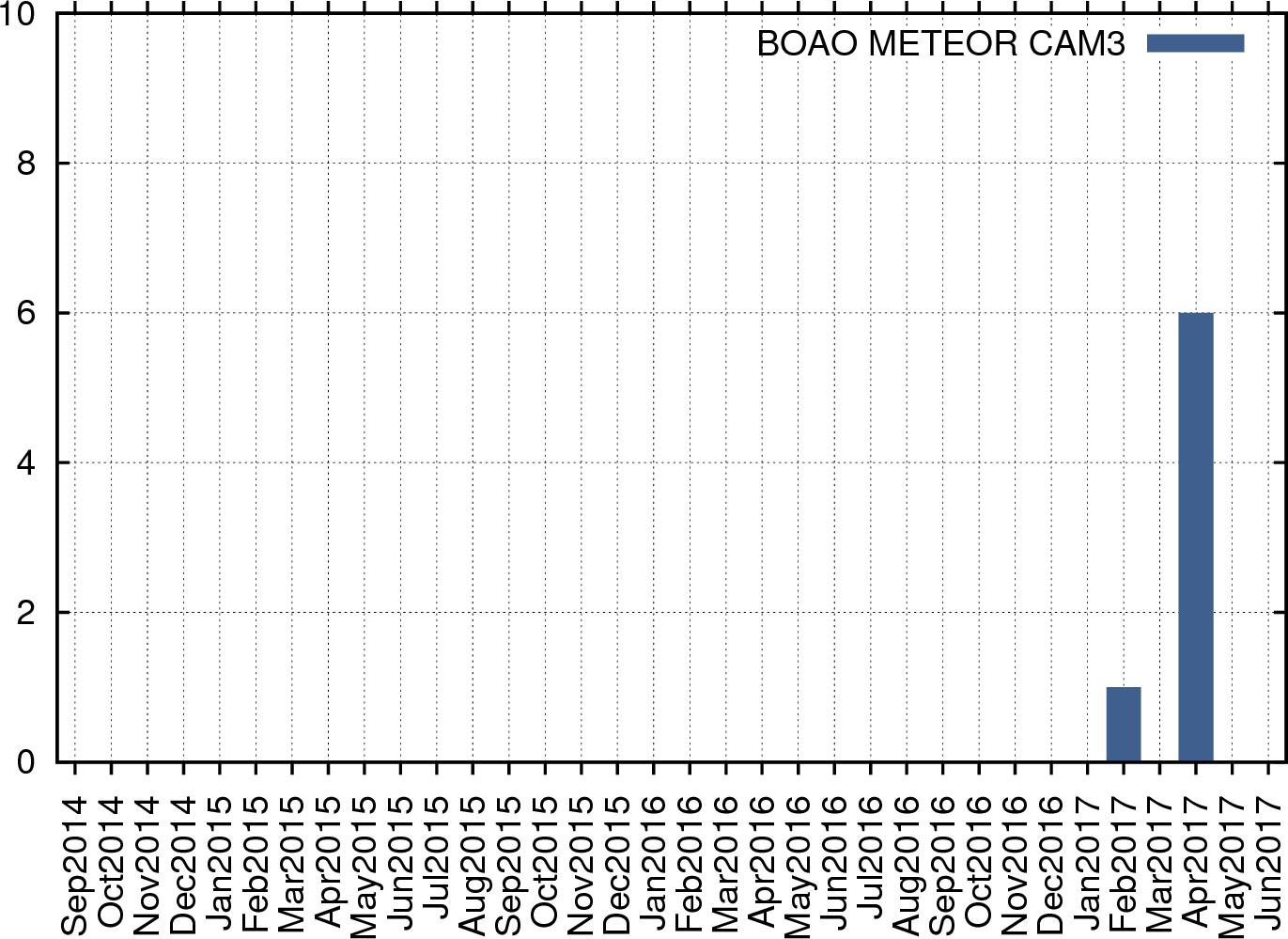}
}}
\caption{Detection statistics for each camera at SOAO (left panels) and 
BOAO (right panels) stations. A total of 1145 and 741 events were 
detected at SOAO and BOAO, respectively. \emph{See online version for colors}.}
\label{histogram1}
\end{figure*}

\subsection{General detection statistics \label{sec:results1}}

The first result concerns the detection statistics at each site and for each camera. For each month we have manually examined the nightly data and recorded the number of meteor detections. Fig.~\ref{histogram1} shows histogram plots of the number of monthly detections for each camera. The recording of meteor events at SOAO stars in October 2014 while detections made at BOAO starts one month later. For the remaining period of 2014 all cameras except for camera 3 at BOAO recorded valid meteor events including double-station detections. We remind the reader about the different scales used on the secondary axis. The detection histogram shows several features.

First we note a lack of detections during several months. This is explained by periods of non-operations of the network mainly due to technical problems with the PC. In particular in June 2015 a lightening strike damaged the SOAO PC permanently due to lack of the installation of a surge-protection-device. We were lucky that such a device was installed at the camera end. As a pre-cautionary consequence, we therefore disconnected all video cables at both SOAO and BOAO observatories to avoid further damage for the remaining summer of 2015. This explains the pausity of detections from June 2015 until around September/October 2015 after which we re-attached the video cables at BOAO and SOAO. With the current use of a coaxial cable we now routinely disconnect all cables from middle of May till middle of September to avoid lightening damage. For the remaining months we experienced additional problems related to faulty 
12V power supply at BOAO and wrong iris and/or software settings at SOAO. Those problems were solved as of today, but still a few remain unsolved. Despite those problems and since first-light of each camera, we recorded a total of 1145 meteor triggers at SOAO and 741 at BOAO. The grand total sums up to 1886.

\begin{figure}
\centerline{\frame{\includegraphics[scale=0.37]{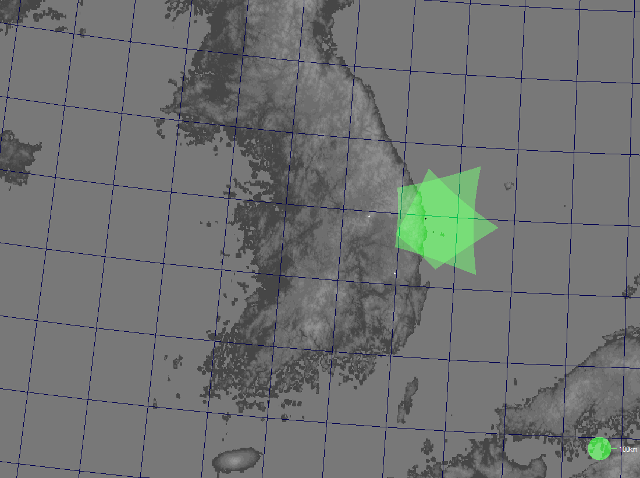}}}
\vspace{1mm}
\centerline{\frame{\includegraphics[scale=0.37]{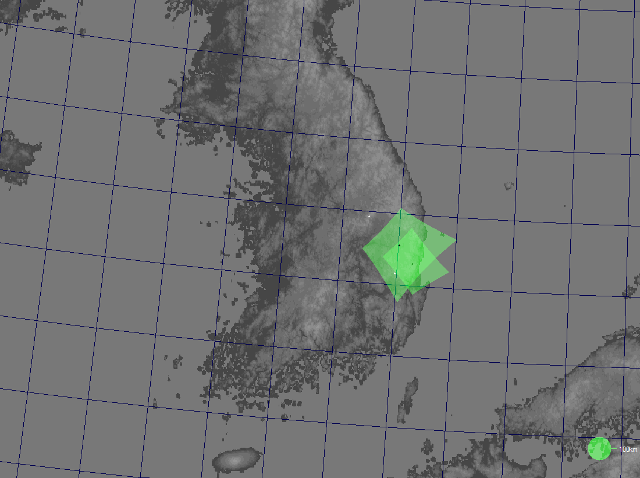}}}
\vspace{1mm}
\centerline{\frame{\includegraphics[scale=0.37]{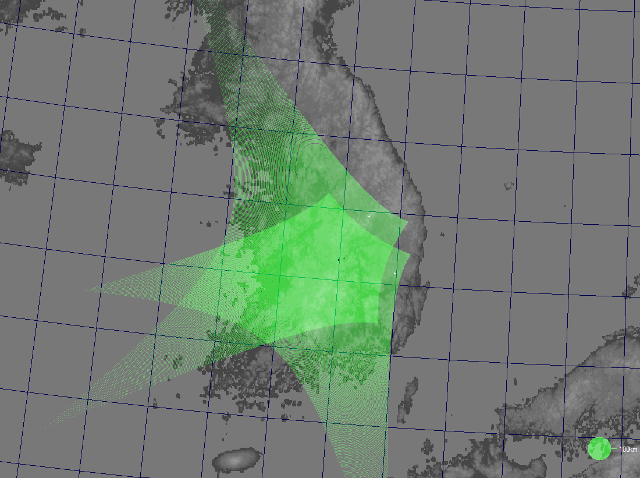}}}
\caption{Ground map plots for pairs of cameras chosing an average meteor height of 100 km. \emph{Top panel:} SOAO/BOAO Cam1. 
\emph{Middel panel:} SOAO/BOAO Cam2. \emph{Bottom panel:} SOAO/BOAO Cam3. Currently the optimum configuration is the Cam1 
pair with similar field of views and large common overlap on the sky. Camera 3 at each station as the largest field of view. 
\emph{See online version for colors}.}
\label{groundmaps1}
\end{figure}

\begin{figure}
\centerline{\includegraphics[scale=0.70]{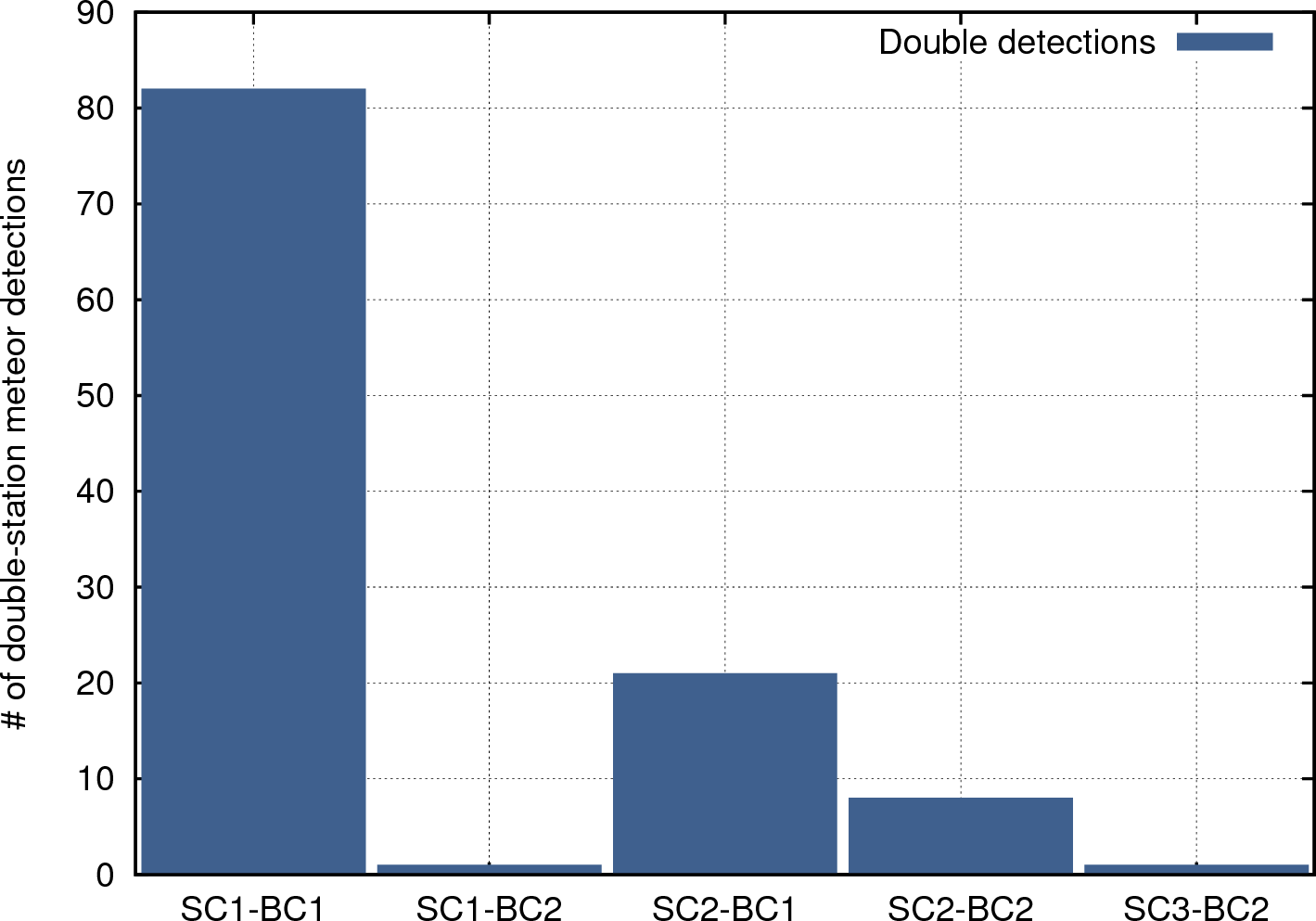}}
\caption{Statistics of double-station meteor detections for possible pairs of cameras. Example: SC1-BC1 stands for SOAOCamera1-BOAOCamera1-pair. \emph{See online version for colors}. As of June 2017 a total of 113 double-station meteors were detected.}
\label{doublestats}
\end{figure}

Second we notice a peak count of meteors in the month of October to December. Especially this can be seen for camera 1,2,3 at SOAO and camera 1 and 2 at BOAO. This increase in meteor detection is likely explained by the observation of the Orionids (October), Leonids (November) and Geminids (December) meteor showers. A detailed radiant determination analysis of all the available data is left for a future study. 

A third feature to notice in Fig.~\ref{histogram1} is the difference scale on the vertical axis. Most detections were made with camera 1 and 2 at both stations with none detections at BOAO camera 3 until February 2017. This is explained by the different field of views. We refer to Table \ref{astrodetails} for details. The larger the field of view the lower the camera is sensitive to detections since the amount of light-flux emitted by the meteor is distributed on a larger field resulting in less flux per pixel. This is mainly the reason why satellite are not detected due to their low intrinsic brightness. Since the field of views of camera 1 and 2 and both ground stations are comparable and significantly smaller than the field of view of camera 3 the result matches our expectations. We conclude that the meteor detection rate efficiency is higher for small field of view lens settings. Further, by comparing the integrated total number of detections of camera 3 at both stations we note that SOAO has a higher detection rate than 
at BOAO. This difference could be explained by a slightly smaller field of view for the SOAO camera 3 lens and fits the general trend for the remaining lens focal length settings. However, part of the difference is explained due to a faulty power-supply unit for BOAO camera 3 during the period June to September 2016.

\subsection{Detection statistics of double-detections\label{sec:results2}}

The main purpose of the meteor cameras is to record double-station meteor events. A double-detection enables the determination of the kinematic and trajectory properties of the meteor. The UAV2 suite allows the plotting of each cameras field of view geometry projected on the ground. The necessary requirement for this task is to perform a astrometric measurement of each cameras orientation and hardware properties. In Fig.~\ref{groundmaps1} we show three combinations of camera pairs. A total of 113 double-station meteors were detected. 

In the following we present qualitative resuls. The top figure panel shows camera pair 1, middle panel shows pair 2 and the lower panel shows pair 3 at each site. A larger overlap is directly related to a higher probability of detecting a double-station event. In agreement with the astrometric measurements the projected field of views are smaller for pair 1 and 2 with the largest projected sky-coverage for pair 3. Relatively, the most optimal camera orientation is for pair 1 displaying the largest percentage coverage of air-space towards eastern direction. The orientation of camera pair 2 display also some overlap. Here, we see some potential for optimisation. By increasing the field of view of camera 2 at BOAO the percentage overlap could be increased and hence a higher double-detection rate is expected. For the camera 3 pair we also note the potential of optimisation of field coverage by change in camera 
orientation. However, as discussed earlier, the larger field of view of camera 3 is not in favour of a higher detection rate. Even a perfectly oriented camera pair would not increase the probability of a double-station detection. The most decisive factor is the camera field of view. 

In Fig.~\ref{groundmaps1} we do not show overlap geometries for additional camera pairings. We will call these pairing non-normal. However, some sky-portion overlap exist for camera 1 (SOAO) and camera 2 (BOAO) as well as camera 2 (SOAO) and camera 1 (BOAO) and also camera 3 (SOAO) and camera 2 (BOAO).

Statistics of double-station detection observations is shown in Fig.~\ref{doublestats}. We report raw numbers. As a note of caution the numbers are to some extent biased as a consequence of the various technical issues that we faced resulting in the non-operation of one station over the other. However, most detections were made during the first 8 to 9 months during which the system did not experience any major issues. During this initial period the detections made are done on a fair ground between the two stations. Therefore, to some qualitative confidence the numbers reported reflect a close to true counting 
statistics. 

The differences in camera geometry and orientation turns out to allow us to conclude a few important aspects of setting up a meteor camera system. Fig.~\ref{doublestats} shows various pairs of cameras. For example SC1-BC1 stands for camera pair 1 at SOAO and BOAO and displays the highest detection rate of double-station meteor events. This is in good agreement with our previous qualitative findings and confirms that not only a small field of view increases detection efficience, but also the overlapping percentage of sky-coverage is important since the camera pair 2 (SC2-BC2) has a similar-size field of view. The detection rate for pair is significantly smaller by a factor of about 8. Surprisingly, we find double-detections for non-normal camera pairs. The non-normal pair SC2-BC1 displays a factor of 2 higher counting rate than the normal pair SC2-BC2 and about 20 times higher than the SC1-BS2 pair.

\begin{figure}[!h]
\centerline{\frame{\includegraphics[scale=0.19]{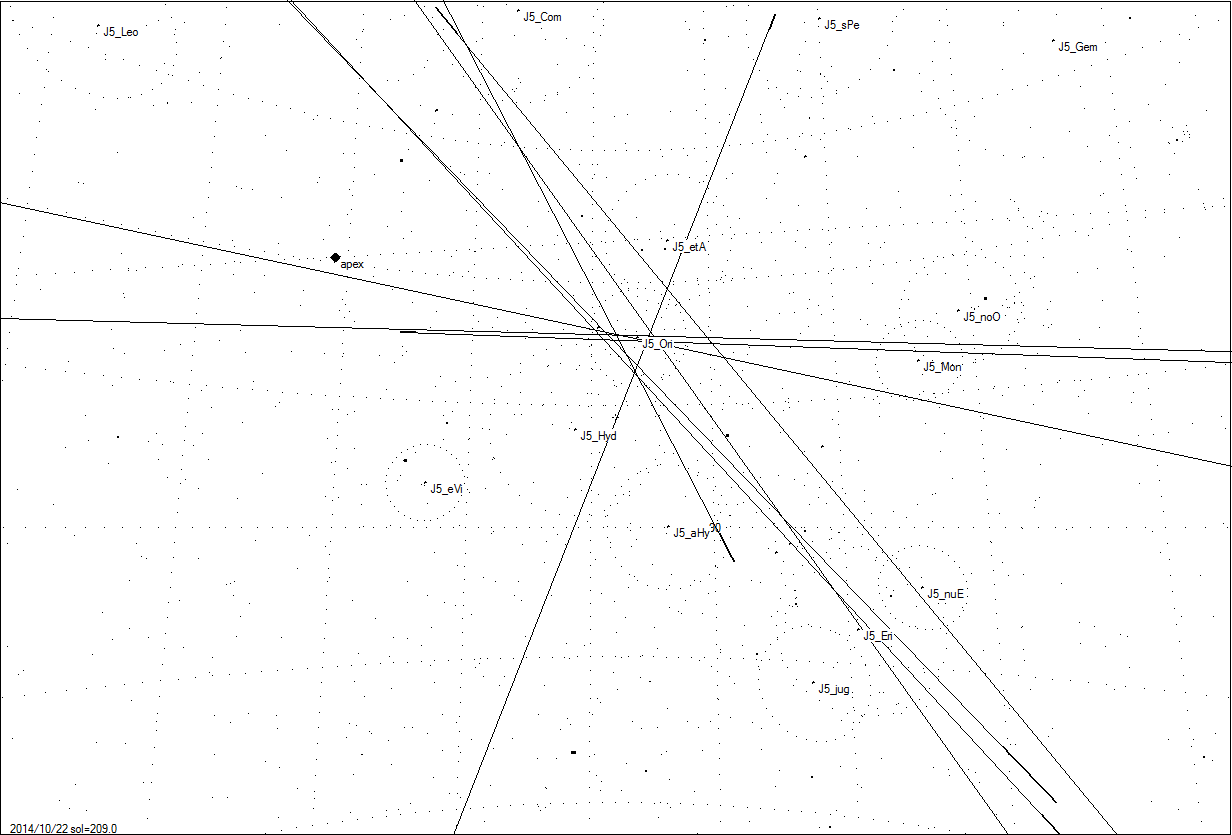}}}
\caption{Radiant map of 9 Orionids as recorded by SOAO camera 1 (four in count) and 3 (five in count). The map employes a gnomonic projection for which a great circle appears as a straight line. Meteor trails appear as straight lines and hence are great circles. The radiant point is located just to the north-east of the Orion shoulder. None are double 
detections. Their classification was carried out with UAV2 based on an assumed atmospheric entry height of 100 km. \emph{See online version for colors}.}
\label{orionradiant}
\end{figure}

\begin{figure}[!h]
\centerline{\frame{\includegraphics[scale=0.24]{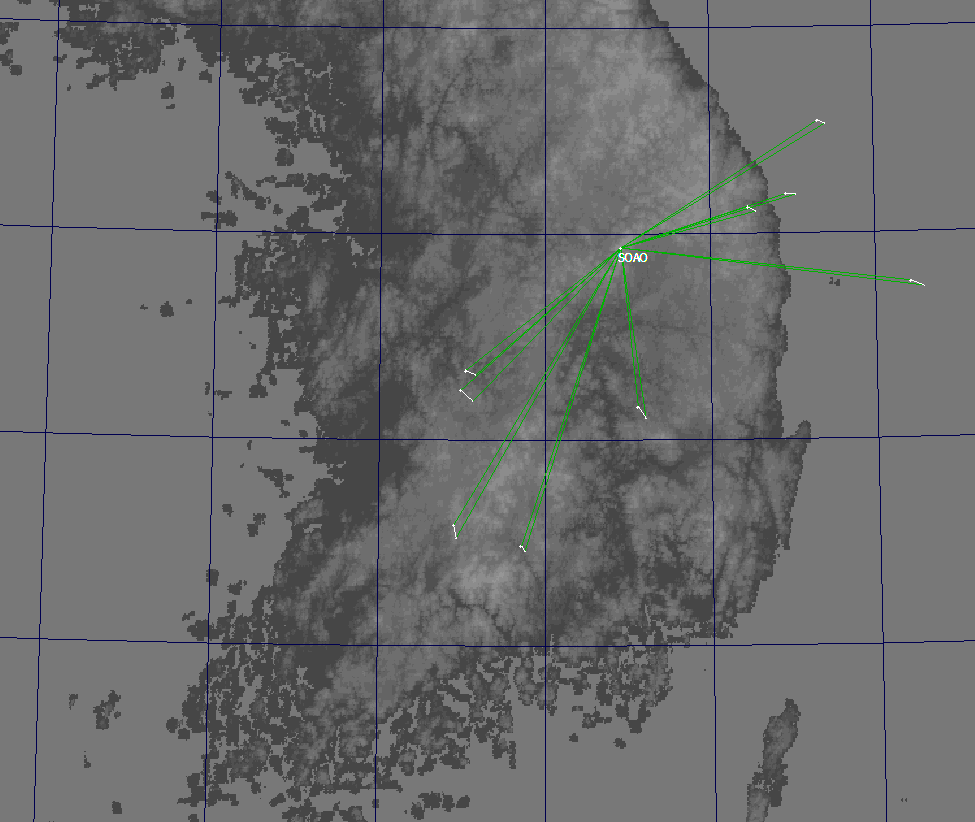}}}
\caption{Ground-trail (projected trails) map of meteors shown in Fig.~\ref{orionradiant} based on an assumed height of 100 km. Meteor events towards the east were recorded with camera 1 (early evening) and trails towards south-western direction were recorded with camera 3 (early morning) at SOAO. \emph{See online version for colors}.}
\label{ground}
\end{figure}

An interesting point to notice in Fig.~\ref{doublestats} is orientation of camera 3 at SOAO and BOAO. The Jinju fireball would have been detected and its orbit tracked easily by the current camera setup. We refer the reader to Fig.~\ref{soao_boao_jinjupath}.

\subsection{Radiant of Orionid meteor shower}

In the later analysis we noted that the orientation of cameras 1 and 3 at SOAO captured the rise and setting of the constellation of Orion. We refer the reader to the lower right panel of Fig.~\ref{variousdetections} for a double-meteor detection of two Orionid meteors using camera 3 at SOAO. We have therefore compiled data for several potential originators from the Orionids meteor shower and determined the corresponding radiant point. The shower radiant is the point on the sky from which members of a meteor shower appears to originate from. The associated meteors could in principle appear anywhere on the sky. By tracing out their meteor trails in revers the radiant point can be determined and hence a possibly new meteor shower identified. The result is shown in Fig.~\ref{orionradiant} and agrees well with the radiant point listed in the general meteor shower catalog of the International Meteor Organisation. The Orionid radiant point is located to the north-western direction from the top-left shoulder of the constellation of Orion. We refer to the figure caption for further details. In Fig.~\ref{ground} we plot the ground-map trails of Orionid meteors as detected from from camera 1 (beginning of night) and 3 
(end of night) at SOAO.

\subsection{Analysis of two double-station detections}

\begin{figure}
\centerline{\frame{\includegraphics[scale=0.20]{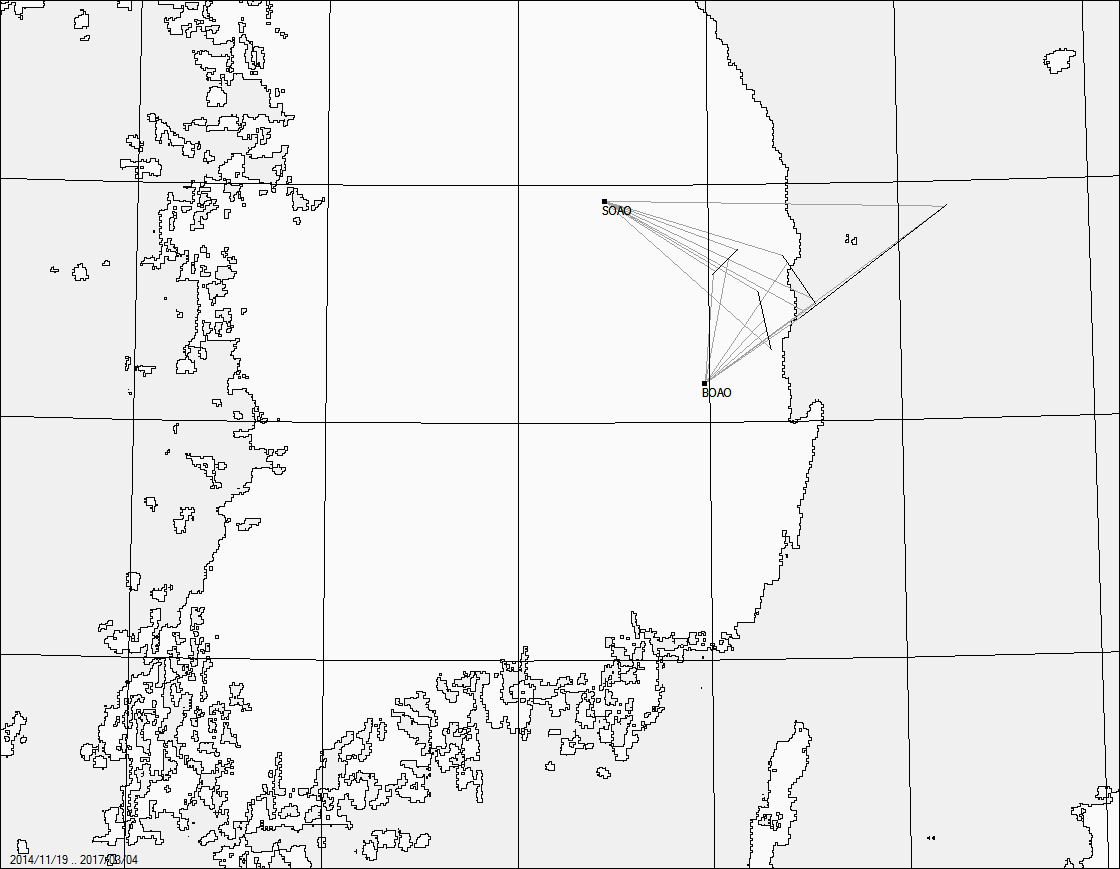}}}
\caption{Four double-station detections. Two meteors were analyzed in some detail. See Table \ref{basictrajparams} and \ref{kinparams} for details on the atmospheric trajectory and kinematics. \emph{See online version for colors}.}
\label{fourdouble}
\end{figure}

We analyzed in some detail two double-station meteors. In the solar system the source region for meteors is either a progenitor body of asteroidal or of cometary origin. We used the result from UAV2 to detect double-station meteors within the UOV2 package. The two meteors were detected immediatedly and paired as a single meteor event. One meteor was observed to travel from an almost northerly direction (azimuth angle of almost 0 degrees) while the second meteor entered the atmosphere from an almost eastern direction (azimuth around 90 degrees). The latter would have been easily observed from the two dokdo islands off the eastern coast of South Korea. We show the two projected meteor trails in a ground-map in Fig.~\ref{fourdouble} alongside with two additional 
double-station meteor detections.

The orbit analyser software UOV2 allows the determination of basic atmospheric trajectory parameters: the equatorial longitude and latitude as well as the site-meteor range and height of the beginning and end of the observed meteor trail. The length of the trail is also estimated. These information provide the meteor atmospheric trajectory. From the frame time sampling, event duration and the assumed linear on-sky meteor trail the three-dimensional velocity vector is determined. We display basic information on the atmospheric trajectory and kinematic properties in Table \ref{basictrajparams} and \ref{kinparams}. In both cases the atmospheric entry height is around 100 km. The travel distance for the asteroidal type was just under 19 km and for the cometary type meteor was around 89 km. The pre-atmospheric entry velocity ($v_{\infty}$) is determined from extrapolation from the measured velocity $v_{obs}$ relative to a given ground-station (SOAO/BOAO). For the cometary meteor event $v_{\infty}$ was around 67 km/s and for the asteroidal meteor event around 31 km/s. These speeds are typical for the two types of meteors and is also reflected in their respective difference in orbital eccentricity. However, UOV2 assumes that no atmospheric drag alters the measured speed. In Table \ref{kinparams} we also provide details on the entry speeds speeds relative to 
Earth $v_{geo}$ and the heliocentric speed $(v_{hel})$. Corrections of the apparent motion of the Earth and the Sun are applied accordingly. The angular velocity ($v_{\theta}$) is depending on the specific geometry and applies only for the vantage-point from a specific camera. 

In Table \ref{radiantparams} we give details on the radiant point calculations. The measured location $(\alpha_{obs},\delta_{obs})$ is called the apparent radiant point and is displaced by two effects \cite{wylie1939} which need to be corrected for: i) \emph{diurnal aberration} (displacement by the rotation of the Earth) and ii) \emph{zenith attraction} (displacement due to the gravitational attraction of the Earth) which is a correction to the apparent velocity of the meteor. The resulting corrected or modified radiant point is listed in Table \ref{radiantparams} in the equatorial as well as ecliptic coordinate system. We found the asteroidal meteor event to be of sporadic nature and was likely transported to the inner realm of the Solar System by repetitive jumps in eccentricity due to orbital resonances with one or more larger planet. The cometary event was classified as belonging to the J5\_Com meteor-stream and is possibly generated by comet Com652 which was discovered by Bourvard in 1798. In Table \ref{orbitparams} we show the calculated orbital parameters of the two events. The marked differences is the semi-major axis (or orbital period) and eccentricity of the two orbits. Both orbits had a perihelion distance of less than one astronomical unit rendering both of them to be crossing the orbit of Earth.

We note that in Table \ref{basictrajparams}, \ref{kinparams}, \ref{radiantparams} and \ref{orbitparams} we do not quote any form of parameter uncertainty. The reason for this is the lack of this information from the UOV2 package. The role of parameter uncertainties and their determination seems to be addressed in \cite{sonotaco2016}. It is possible that a future release of UOV2 will also include error propagation from video meteor measurements to determine uncertainties.

\begin{table*}
\begin{center}
\centering
\caption{Atmospheric trajectory parameters (topocentric equatorial system) of two meteor events. Parameter explanation - values in () are used variable designation in UOV2:
S - start of trail (1). E - end of trail (2). $\lambda$ - azimuth angle (az). $\phi$ - altitude angle (ev). $R$ - range from ground station to meteor (LD). $H$ - height of meteor above ground (H). $L$ - length of meteor trail (LD21). The MJD time stamp is valid at the beginning of event. For parameter uncertainties we refer the reader to the main text.}
\begin{tabular}{ccccc} 
\hline
~                                      & SOAO Cam1 & BOAO Cam1                   & SOAO Cam1 & BOAO Cam1            \\
~                                      & \multicolumn{2}{c}{1: cometary}         & \multicolumn{2}{c}{2: asteroidal}\\
\hline
\hline
Event time stamp (UTC)                       & \multicolumn{2}{c}{20161128\_133033}    & \multicolumn{2}{c}{20170304\_155304} \\
MJD (UTC)                              & \multicolumn{2}{c}{57720.56289}         & \multicolumn{2}{c}{57816.88010} \\
(J2000.0) $\lambda_S$~(deg.)           &      91.072    &            54.269      & 110.502   & 11.398               \\      
(J2000.0) $\lambda_E$~(deg.)           &      118.960   &            56.259      & 122.576   & 5.030                \\      
(J2000.0) $\phi_S$~(deg.)              &      34.271    &            38.005      & 53.585    & 54.766               \\      
(J2000.0) $\phi_E$~(deg.)              &      40.784    &            59.724      & 51.958    & 55.930               \\      

$R_S$ (km)                             &      196.875   &            181.504     & 113.699   & 103.628              \\      
$R_E$ (km)                             &      145.381   &            110.304     & 101.392   & 92.531               \\

$H_S$ (km)                             & \multicolumn{2}{c}{115.024}             & \multicolumn{2}{c}{90.726}       \\
$H_E$ (km)                             & \multicolumn{2}{c}{96.874}              & \multicolumn{2}{c}{79.013}       \\
$L$   (km)                             & \multicolumn{2}{c}{89.005}              & \multicolumn{2}{c}{18.637}       \\
\hline
\end{tabular}
\label{basictrajparams}
\end{center}
\end{table*}

\begin{table*}
\begin{center}
\centering
\caption{Basic kinematic data (magnitudes) of two meteor events (same as in Table \ref{basictrajparams}). Parameter explanation - values in () are used variable designation in UOV2:
$v_{obs}$ - observed relative velocity between object and station (vo). $v_\infty$ - extrapolated pre-atmospheric velocity or initial entry speed (vi). 
$v_{geo}$ - geocentric velocity at entry (vg). $v_{hel}$ - heliocentric velocity at entry. $v_{\theta}$ - average angular velocity at each station (va). For parameter uncertainties we refer the reader to the main text.}
\begin{tabular}{c c c c c}
\hline
~                                      & SOAO Cam1 & BOAO Cam1                     & SOAO Cam1 & BOAO Cam1            \\
~                                      & \multicolumn{2}{c}{1: cometary}           & \multicolumn{2}{c}{2: asteroidal}\\
\hline
\hline
Event time stamp (UTC)                 & \multicolumn{2}{c}{20161128\_133033}      & \multicolumn{2}{c}{20170304\_155304} \\
MJD (UTC)                              & \multicolumn{2}{c}{57720.56289}           & \multicolumn{2}{c}{57816.88010}  \\

event duration (s)                     & \multicolumn{2}{c}{1.335}                 & \multicolumn{2}{c}{0.584} \\
$v_{obs}$ (km/s)                       & \multicolumn{2}{c}{66.8}                  & \multicolumn{2}{c}{30.7}  \\
$v_{\infty}$ (km/s)                    & \multicolumn{2}{c}{66.8}                  & \multicolumn{2}{c}{30.7}  \\
$v_{geo}$ (km/s)                       & \multicolumn{2}{c}{65.5}                  & \multicolumn{2}{c}{28.4}  \\
$v_{hel}$ (km/s)                       & \multicolumn{2}{c}{42.1}                  & \multicolumn{2}{c}{36.2}  \\
$v_{\theta}$ (deg./s)                  & 17.3                      & 14.1          & 13.0                       & 8.9  \\

\hline
\end{tabular}
\label{kinparams}
\end{center}
\end{table*}

\begin{table*}
\begin{center}
\centering
\caption{Measured radiant parameters for two meteor events (same as in Table \ref{basictrajparams}). Parameter explanations - values in () are used variable
designation in UOV2: $\alpha, \delta$ - observed and modified right ascension and declination of radiant point in the equatorial system. (ra\_o, dc\_o,ra\_t,dc\_t). R.A, Dec. - right ascension and declination (J2000.0) of modified radiant point in the ecliptic system (elng, elat). For parameter uncertainties we refer the reader to the main text.}
\begin{tabular}{ccccc}
\hline
~                                      & SOAO Cam1 & BOAO Cam1                     & SOAO Cam1 & BOAO Cam1            \\
~                                      & \multicolumn{2}{c}{1: cometary}           & \multicolumn{2}{c}{2: asteroidal}\\
\hline
\hline
Event time stamp (UTC)                 & \multicolumn{2}{c}{20161128\_133033}      & \multicolumn{2}{c}{20170304\_155304} \\
$\alpha_{obs}$ (J2000.0)~(deg.)        & \multicolumn{2}{c}{143.48840}             & \multicolumn{2}{c}{241.88009} \\
$\delta_{obs}$ (J2000.0)~(deg.)        & \multicolumn{2}{c}{36.004400}             & \multicolumn{2}{c}{53.601120} \\
$\alpha_{mod}$ (J2000.0)~(deg.)        & \multicolumn{2}{c}{144.23801}             & \multicolumn{2}{c}{244.74594} \\
$\delta_{mod}$ (J2000.0)~(deg.)        & \multicolumn{2}{c}{35.78718}              & \multicolumn{2}{c}{53.73289}  \\
R.A. (J2000.0)~(deg.)                  & \multicolumn{2}{c}{134.59767}             & \multicolumn{2}{c}{213.99347} \\
Dec. (J2000.0)~(deg.)                  & \multicolumn{2}{c}{20.363199}             & \multicolumn{2}{c}{72.278198} \\
\hline
\end{tabular}
\label{radiantparams}
\end{center}
\end{table*}

\begin{table}
\begin{center}
\caption{Derived osculating orbital Keplerian elements at 
pre-atmospheric entry of two meteor events (same as in Table \ref{basictrajparams}. Parameter explanation - values in () are 
used variable designations in UOV2: $a$ - semi-major axis (a). $e$ - eccentricity (e). $I$ - orbital inclination (incl). $\omega$ - argument of pericenter (peri). 
$\Omega$ - argument of ascending node (node). $P$ - orbital period (p). $q$ - perihelion distance (q). For parameter uncertainties we refer the reader to the main text.}
\begin{tabular}{lcc}
\hline
parameters            &        cometary orbit    &         asteroidal orbit \\
\hline
\hline
MJD (UTC)             &       57720.56289        &        57816.88010       \\
$a$ (AU)              &           38.6           &           1.8            \\
$e$                   &           0.98           &           0.48           \\
$I$ (deg.)            &           139.6          &           49.2           \\
$\omega$ (deg.)       &           247.6          &           205.4          \\
$\Omega$ (deg.)       &           246.5          &           344.0          \\
$P$ (yrs)             &           240.3          &           2.5            \\
$q$ (AU)              &           0.68           &           0.96           \\
\hline
\end{tabular}
\label{orbitparams}
\end{center}
\end{table}

\section{Summary and future plans \label{sec:conclude}}

This paper has given a detailed description of the planning, installation and operation of a double-station video-meteor detection system in the Republic of Korea. 
The level of details given was chosen to be relatively high for two purposes: i) currently no comparable system exist in South Korea and hence 
would need detailed documentation and ii) in order to allow future amateur-astronomers in the Republic of Korea (or elsewhere in 
the world) to install their own detection system, this paper would provide the necessary technical background information and might even provide motivation for the installation of such a system. The project was initiated around June 2014 within the frame-work of a Research \& Education programme between Daejeon Science Highschool and KASI. The first successful 
detection of an atmospheric double-station meteor event was recorded a few month later on November 18, 2014 from SOAO (camera 1) and BOAO (camera 2). The most recent double-station meteor detection was on March 4, 2017. Many meteor events were recorded in between this time period and the system is currently still maintained and operating. We have therefore succesfully demonstrated a working example of a professional setup for non-stop 
nightly meteor detection in South Korea. However, the present system is not optimized. The installation of the proto-type system has helped to gain valuable experience which 
will benefit future installations of similar but optimized systems. For the future we plan to address the following action-items:

\begin{itemize}
\item Installation of a 7th camera at SOAO with a blazed grating mounted infront of the lens for spectroscopic measurements.
\item Replacement of coaxial cable with a fibre-optics cable (including a multi-plexer video converter) to avoid damage to electronic equpment due to lightening strikes. This would also enable the operation of the cameras during the summer period during the months of June, July and August.
\item We plan to install additional hard-drives in each PC. This ensures that each capture device can write data over an independent bus and hence potentially decreases the risk of frame-drops.
\item In the fall of 2017 we plan to carry out test towards localising the blue-screen-of-death problem at SOAO PC by increasing cooling air-circulation.
\item To improve timing precision we plan to install GPS based timing equipment allowing timing precision on the order of $10^{-4}$ seconds.
\item From the ground map analysis we found that the camera orientation has to be re-adjusted in order to maximise overlapping-percentage of the respective field of views. This is particulary necessary for SOAO/BOAO-camera-3 and SOAO/BOAO-camera-2 pair.

\item In the more far future we plan to install an array of cameras each having a small field of view (30 x 20 deg.) to ensure all-sky monitoring. A smaller field of view enables higher astrometric accuracy and the detection of fainter meteors.
\item We also plan to test for single-camera all-sky cameras using fish-eye lenses.

\item We also plan, in collaboration with Dr. David Asher (Armagh Observatory), to assess wether solar system planets will be moving in the field of view of any of the six cameras. If that is the case, then the positions of such planets could be recorded over a time scale of months and their orbits calculated as part of a Research \& Education public outreach project.

\item We plan to carry out a future statistical analysis of SOAO/BOAO meteor data using R programming language (for more information see: http://meteornews.org/r-suite-analysis-edmond-database/)

\end{itemize}

\acknowledgments

The realisation and completion of this project would not have been possible without the help of numerous individuals who were willing to help out in one or the other way.
In particular we would like to thank Dr. Hongsu Kim (KASI) and Dr. Daeho Yoo (then at KAIST) for their assistance requested at any time. Further the help of Mr. Dak-Hyon Cho (New Continent Engineering Co., Daejeon, Republic of Korea) and his team was of great importance helping and partially sponsoring the installation of the equipment at SOAO and BOAO in the fall of 2014. Also we would like to thank Mr. Noh (KASI) and his team for his assistance with planning the installation. Additionally, KASI is acknowledged to have sponsored computing equipment and personal to ensure the correct installation of cables and power-supply. Furthermore, we would like to thank the staff at the SOAO/BOAO observatory for helping with solving on-site problems on short notice: Mr. Yoon-Ho Park, Sang-Min Lee, Hyun-Il Sung, Hyung-Il Oh, Jang-Ho Park, Taek-Su Kim, Eon-Chang Sung. Further we would like to thank for answering questions via email and for providing solutions to technical difficulties: Drs. Apostolos Christou, Detlef Koschny, Sirko Molau, David Asher, William Stuart. We are sure there are other people who have helped out in one or the other way and we would like to apologize for not naming them here at this stage. If you would like to be mentioned on the official project home page (http://meteor.kasi.re.kr), then please contact the lead author of this manuscript.

\section{Appendix}

\subsection{Technical mount specifications}

\begin{figure*}
\centerline{\includegraphics[scale=0.7,angle=-90]{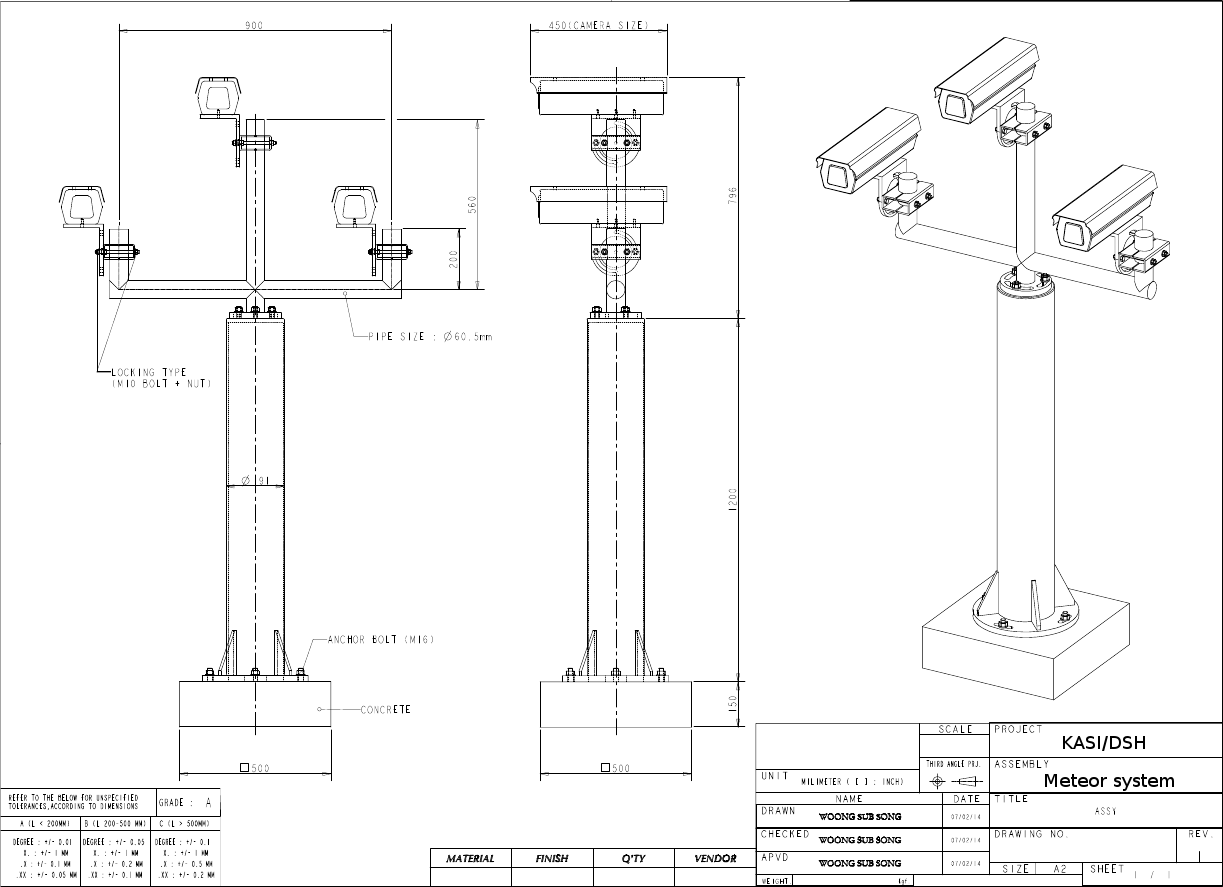}}
\caption{Technical data and dimensions for the mounting and camera system.}
\end{figure*}


\begin{thebibliography}{}

\bibitem[Ahn(2002)]{ahn2002}
Ahn, I. 2002, MSc. thesis, Seoul National University, South Korea

\bibitem[Ahn(2003)]{ahn2003}
Ahn, S.-H., 2003, MNRAS, 343, 1095-1100

\bibitem[Ahn(2004)]{Ahn2004}
Ahn, S.-H., 2004, Earth Moon and Planets, 95, 63-68

\bibitem[Ahn(2005)]{ahn2005}
Ahn, S.-H., 2005, MNRAS, 358, 1105-1115

\bibitem[Ahn(2015)]{ahn2015}
Ahn, S.-H., 2015, Proceedings of the IMC, Mistelbach 2015, p.84

\bibitem[Ahn(2016)]{ahn2016}
Ahn, S.-H., 2016, Programme \& Abstract Booklet for Meteoroids 2016 Conference, 6-10 June 2016, ESA/ESTEC, Noordwijk, The Netherlands, 2016

\bibitem[Atreya \& Christou(2007)]{atreyaIDL}
Atreya, P. \& Christou, T., 2007, Proceedings of the International Meteor Conference, Roden, The Netherlands, 14-17 September, 2006 Eds.: Bettonvil, F., Kac, J. International Meteor Organization, ISBN 2-87355-018-9, pp.18-23

\bibitem[Atreya et al.(2011)]{atreya2011}
Atreya, P., Vaubaillon, J., Colas, F., Bouley, S., Gaillard, B., Sauli, I., Kwon, M.-K., 2011, Proceedings of the International Meteor Conference, Armagh, Northern Ireland, 16-19 September, 2010 Eds.: Asher, D.J.; Christou, A.A.; Atreya, P.; and Barentsen, G. International Meteor Organization, ISBN 2978-2-87355-022-6, pp. 10-13

\bibitem[Brown et al.(2010)]{asgard2010}
Brown, P., Weryk, R. J., Kohut, S., Edwards, W. N., Krzeminski, Z., 2010, WGN, Journal of the International Meteor Organization, vol. 38, no. 1, p. 25-30

\bibitem[Blaauw \& Cruse(2012)]{blaauw2012}
Blaauw, R.; Cruse, K. S., 2012, Proceedings of the International Meteor Conference, Sibiu, Romania, 15-18 September, 2011 Eds.: Gyssens, M.; and Roggemans, P. International Meteor Organization, ISBN 2978-2-87355-023-3, pp. 44-46

\bibitem[Choi, Lee, \& Shin(2002)]{choi2002}
Choi, B. -G., Ahn, I., Lee, M. S. and Shin H. -S. 2002, Geosciences Journal, 6, 161-167

\bibitem[Christou \& Atreya(2007)]{armaghmeteor2007}
Christou, T., \& Atreya, P., 2007, Proceedings of the International Meteor Conference, Roden, The Netherlands, 14-17 September, 2006 Eds.: Bettonvil, F., Kac, J. International Meteor Organization, ISBN 2-87355-018-9, pp.141-145

\bibitem[Campbell-Burns \& Kacerek(2014)]{ukmon2014}
Campbell-Burns, P., Kacerek, R., 2014, WGN, Journal of the International Meteor Organization, vol. 42, no. 4, p. 139-144

\bibitem[Choi et al.(2014)]{choi2014}	
Choi, Y., Kim, M., Byun, Y., Yi, H., Chang, S., Choi, J., Sohn, J., Moon, H., Park, J., 2014, Asteroids, Comets, Meteors 2014. Proceedings of the conference held 30 June - 4 July, 2014 in Helsinki, Finland. Edited by K. Muinonen et al.

\bibitem[Colas et al.(2015)]{colas2015}
Colas, F., Zanda, B.; Vaubaillon, J. et al., 2015, Proceedings of the International Meteor Conference, Mistelbach, Austria, 27-30 August 2015, Eds.: Rault, J.-L.; Roggemans, P., International Meteor Organization, ISBN 978-2-87355-029-5, pp. 37-40

\bibitem[Choi et al.(2015)]{choi2015}
Choi, B.-G., Kim, H., Kim, H., Lee, J. I., Kim, T. H., Ahn, I., Yi, K., Hong, T. E., 2015, 78th Annual Meeting of the Meteoritical Society, July 27-31, 2015. LPI Contribution No. 1856, p. 5091

\bibitem[Fleet(2015)]{fleet2015}
Fleet, R., 2015, Proceedings of the International Meteor Conference, Mistelbach, Austria, 27-30 August 2015, Eds.: Rault, J.-L.; Roggemans, P., International Meteor Organization, ISBN 978-2-87355-029-5, pp. 30-32

\bibitem[Grady(2000)]{grady2000}
Grady, M. M. 2000, Catalogue of Meteorites. 5th ed. Cambridge University Press, Cambridge, p. 689

\bibitem[Goh \& Choi (2016)]{goh2016}	
Goh, S., Choi, B.-G., 2016, 79th Annual Meeting of the Meteoritical Society, 7-12 August, 2016. LPI Contribution No. 1921, p. 6270

\bibitem[Gural(2008)]{gural2008}
Gural, P. S., 2008, EM\&P, vol. 102, issue 1-4, p. 269-275

\bibitem[Jenniskens et al.(2011)]{CAMS1}
Jenniskens, P. et al. 2011, "CAMS: Cameras for Allsky Meteor Surveillance to establish minor meteor showers", Icarus, 216, 40

\bibitem[Jenniskens(2017)]{jen2017}
Jenniskens, P., 2017, P \& SS, 143, 116

\bibitem[Kim et al.(1145)]{kim1145}
Kim, B.-S. et al., 1145, Goryeo, [{\it Samguksagi}]

\bibitem[Kim et al.(1451)]{kim1451}
Kim, J.-S. et al., 1451, Joseon, [{\it Goryeosa}]

\bibitem[Kim et al.(1452)]{kim1452}
Kim, J.-S. et al., 1452, Joseon [{\it GoryeosaJoryo}]

\bibitem[Kanamori et al.(2009)]{kanamori2009}
Kanamori, T., Int, J., Jenniskens, P., Jopek, T., 2009, Central Bureau Electronic Telegrams, No. 1771, \# 1 (2009). Edited by Green, D. W. E.

\bibitem[Koschny et al.(2014)]{koschny2014}
Koschny, D., Mc Auliffe, J., Drolshagen, E., 2014, Proceedings of the International Meteor Conference, Giron, France, 18-21 September 2014 Eds.: Rault, J.-L.; Roggemans, P. International Meteor Organization, ISBN 978-2-87355-028-8, pp. 10-15

\bibitem[Molau(1994)]{molau1994}
Molau, S., 1993, Proceedings of the International Meteor Conference, Puimichel, France, 
23-26 September 1993, Eds.: Roggemans, P., International Meteor Organization, p. 71-75

\bibitem[Montenbruck \& Pfleger(2000)]{montenbruck}
Montenbruck,O. \& Pfleger, T., 2000, ``Astronomy on the Personal Computer'', ISBN: 3-540-67221-4, Springer Verlag, Germany

\bibitem[Nagao et al.(2015)]{nagao2015}	
Nagao, K., Haba, M. K., Lee, J. I., Kim, T., Lee, M. J., 2015, 78th Annual Meeting of the Meteoritical Society, July 27-31, 2015. LPI Contribution No. 1856, p. 5027

\bibitem[Silber(2014)]{silber2014}
Silber, E. A., 2014, Proceedings of the International Meteor Conference, Poznan, Poland, 22-25 August 2013. Eds.: Gyssens, M.; Roggemans, P.; Zoladek, P. International Meteor Organization, ISBN 978-2-87355-025-7, pp. 136-138

\bibitem[Silber et al.(2009)]{silber2009}
Silber, Elizabeth A., ReVelle, Douglas O., Brown, Peter G., Edwards, Wayne N., 2009, Journal of Geophysical Research, Volume 114, Issue E8

\bibitem[SonotaCo et al.(2013)]{UCV2}
SonotaCo, UFOCaptureV2, 2013, URL: http://sonotaco.com/soft/UFO2/help/english/index.html

\bibitem[SonotaCo et al.(2016)]{UAV2}
SonotaCo, UFOAnalyzerV2, 2016, URL: http://sonotaco.com/soft/download/UA2Manual\_EN.pdf

\bibitem[SonotaCo et al.(2016)]{UOV2}
SonotaCo, UFOOrbitV2, 2016, URL: http://sonotaco.com/soft/UO2/UO21Manual\_EN.pdf

\bibitem[SonotaCo(2009)]{sonotaco2009}
SonotaCo, 2009, WGN, Journal of the International Meteor Organization, vol. 37, no. 2, p. 55-62

\bibitem[SonotaCo(2016)]{sonotaco2016}
SonotaCo, 2016, WGN, Journal of the International Meteor Organization, vol. 44, no. 2, p. 42-45

\bibitem[Trigo-Rodr{\'i}gues et al.(2007)]{spmn2007}
Trigo-Rodr{\' i}guez, J. M., Madiedo, J. M., Castro-Tirado, A. J., Ortiz, J. L., Gural, P. S., Llorca, J., Fabregat, J., Vitek, S., Pujols, P., Troughton, B., 2014, 38th Lunar and Planetary Science Conference, (Lunar and Planetary Science XXXVIII), held March 12-16, 2007 in League City, Texas. LPI Contribution No. 1338, p.1584

\bibitem[Trigo-Rodr{\'i}gues et al.(2004)]{spmn2004}
Trigo-Rodr{\'i}guez, J. M., Castro-Tirado, A. J., Llorca, J., Fabregat, J., Mart{\'i}nez, V. J., Reglero, V., Jel{\'i}nek, M., Kub{\'a}nek, P., 
Mateo, T.; Postigo, A. De Ugarte, 2004, Earth, Moon, and Planets, Volume 95, Issue 1-4, p. 553-567

\bibitem[Vida et al.(2016)]{vida2016}
Vida, D., Zubovi{\'c}, D. {\v S}egon, D., Gural, P., Cupec, R.
Proceedings of the International Meteor Conference, Egmond, the Netherlands, 2-5 June 2016, Eds.: Roggemans, A.; Roggemans, P., ISBN 978-2-87355-030-1, pp. 307-318

\bibitem[Wylie(1939)]{wylie1939}
Wylie, C. C., 1939, PA, 47, 425.

\end{thebibliography}
\end{document}